\documentclass[journal=jacsat,manuscript=article]{achemso}

\usepackage[utf8]{inputenc}
\usepackage{setspace}
\usepackage{parskip}
\usepackage{amssymb}
\usepackage{amsmath}
\usepackage{graphicx}
\usepackage[english]{babel}
\usepackage{fancyhdr}
\usepackage[T1]{fontenc}
\usepackage{gensymb}
\usepackage{epstopdf}
\usepackage{siunitx}
\usepackage{graphicx} 
\usepackage{float}
\usepackage{mathtools}
\usepackage{color}
\usepackage{xcolor}

\author{Vaibhav Thakore}
\affiliation{Department of Applied Mathematics, Western University, 1151 Richmond Street, London, Ontario N6A\,5B7, Canada}
\email{vthakore@knights.ucf.edu}

\author{Tapio Ala-Nissila}
\email{tapio.ala-nissila@aalto.fi}
\affiliation{QTF Center of Excellence, Department of Applied Physics, Aalto University School of Science, FIN-00076, Aalto, Espoo, Finland}
\alsoaffiliation{Department of Physics, Brown University, Providence, Rhode Island 02912-1843, USA}
\alsoaffiliation{Interdisciplinary Centre for Mathematical Modelling, Department of Mathematical Sciences, Loughborough University, Loughborough LE11 3TU, UK}

\author{Mikko Karttunen}
\affiliation{Department of Physics and Astronomy, The University of Western Ontario, 1151 Richmond Street, London, Ontario N6A\,3K7, Canada}
\email{mkarttu@uwo.ca}
\alsoaffiliation
{The Centre of Advanced Materials and Biomaterials Research,  The University of Western Ontario, 1151 Richmond Street, London, Ontario, N6A\,5B7, Canada}
\alsoaffiliation{Department of Chemistry,  The University of Western Ontario, 1151 Richmond Street, London, Ontario, N6A\,5B7, Canada}

\title[An \textsf{achemso} demo]
  {Temperature-resilient anapole modes associated with TE polarization in semiconductor nanowires\footnote{ Electronic supplementary information (ESI) available. See DOI:}}

\keywords{Thermoplasmonic response; TE and TM polarization, cylindrical nanowires; Mie resonances; anapole modes; indirect and direct bandgap semiconductors; and, noble metals}

\begin{document}

%





\newpage

\begin{abstract}
 Polarization-dependent scattering anisotropy of cylindrical nanowires has numerous potential applications in, for example, nanoantennas, photothermal therapy, thermophotovoltaics, catalysis, sensing, optical filters and switches. In all these applications, temperature-dependent material properties play an important role and often adversely impact performance depending on the dominance of either radiative or dissipative damping. Here, we employ numerical modeling based on Mie scattering theory to investigate and compare the temperature and polarization-dependent optical anisotropy of metallic (gold, Au) nanowires with indirect (silicon, Si) and direct (gallium arsenide, GaAs) bandgap semiconducting nanowires. Results indicate that plasmonic scattering resonances in semiconductors, within the absorption band, deteriorate with an increase in temperature whereas those  occurring away from the absorption band strengthen as a result of the increase in phononic contribution. Indirect-bandgap thin ($20 \,\mathrm{nm}$) Si nanowires present low absorption efficiencies for both the transverse electric (TE, $E_{\perp}$) and magnetic (TM, $E_{\parallel}$) modes, and high scattering efficiencies for the TM mode at shorter wavelengths making them suitable as highly efficient scatterers. Temperature-resilient higher-order anapole modes with their characteristic high absorption and low scattering efficiencies are also observed in the semiconductor nanowires ($r \! = \! 125 \! - \! 130$ nm) for the TE polarization. Herein, the GaAs nanowires present $3 \! - \! 7$ times greater absorption efficiencies compared to the Si nanowires making them especially suitable for temperature-resilient applications such as scanning near-field optical microscopy (SNOM), localized heating, non-invasive sensing or detection that require strong localization of energy in the near field.          
\end{abstract}

\section*{Introduction}
Polarization-dependent scattering anisotropy offers an important tool, among others, to control incident radiation on the mesoscale using sub-wavelength nano- or microstructures for applications in directional nanoantennas \cite{RN150}, photothermal therapy  \cite{RN31}, thermophotovoltaics \cite{RN56}, nanocatalysis \cite{RN19}, biomedical sensing \cite{RN57}, optical filters and switches \cite{RN191}.
Early investigations in plasmonics focused on the so-called 'epsilon-negative' metals because of their strong plasmonic response driven by the large negative real part of their dielectric permittivity \cite{RN195, RN196, RN193}. However, excessive dissipation in metallic nanostructures especially in the near-infrared (NIR) regime has led to a quest for semiconductor materials with low losses that act as dielectrics in the NIR regime away from their absorption band edge \cite{RN159,RN135,RN62}. Studies on Mie resonances in noble metal and dielectric nanowires have been reported previously \cite{RN199, RN197, RN198}, but thus far there have been no systematic studies on the polarization dependence of Mie resonances in metallic or semiconductor nanowires that would also account for their temperature-dependent dielectric permittivities. 
Furthermore, most attention has been on the generation of dark anapole modes using cylindrical metamaterials characterized by near-zero scattering efficiencies and high absorption efficiencies resulting in significantly enhanced near-field energy \cite{RN203}. Anapole modes have been shown to arise from electrodynamic out-of-phase charge-current distributions that do not radiate or interact with external fields as a result of destructive interference of radiation produced by spatially overlapping and co-excited electric, toroidal or magnetic modes that suppresses scattering in the far field \cite{RN205,RN204}.     
However, the temperature and polarization dependence of anapole modes in single cylindrical nanowires has not been studied thus far.    

Here, we present results for temperature and polarization-dependent Mie resonances in metallic (Au), and indirect (Si) and direct (GaAs) bandgap semiconductor nanowires. We then identify anapole modes in the semiconductor nanowires and study their temperature and polarization dependence. The next section, Theory and Methods, 
briefly outlines the analytic solution to the polarization-dependent scattering from long nanowires and the computation of the temperature-dependent dielectric permittivites for Au, Si and GaAs. 
When discussing the results, we first analyze
polarization and temperature-dependent scattering and absorption efficiencies. This is followed by a discussion of the results in the limit of thin wires, polarization and temperature dependence of anapole modes, and, a comparison of radiative to dissipative damping as a function of temperature and polarization. 

Briefly, our results indicate that plasmonic resonances degrade at elevated temperatures in both Au and semiconductor nanowires (within the absorption band edge) regardless of the polarization of the incident radiation. The indirect bandgap thin Si nanowires present lower absorption efficiencies for both the perpendicular (TE) and parallel (TM) polarizations at short wavelengths in conjunction with high scattering efficiencies for the TM mode making them suitable as highly efficient scatterers. Results also point to the existence of higher-order anapole states in both Si and GaAs nanowires. The anapole modes observed for the TM polarization exhibit a large redshift with increase in temperature while those for the TE polarization are observed to be resilient to temperature changes.   

\section*{Theory and Methods \label{sec:theory}}
\subsection*{Scattering from an infinite cylindrical wire}

The numerical computations for the polarization-dependent absorption ($Q_\mathrm{abs}$) and scattering ($Q_\mathrm{sca}$) efficiencies at different temperatures were carried out using Mie theory for cylindrical nanowires \cite{RN102}. The analytical solution is derived by solving the wave equation for the electric and magnetic fields of a plane electromagnetic wave incident on an infinitely long cylindrical scatterer. The infinite cylindrical scatterer and the non-absorbing surrounding medium (refractive index, $n_\mathrm{m} = 1.5$) are assumed to be linear, isotropic and homogeneous wherein the wave equations for the $\textbf{E}$ and $\textbf{H}$ fields are given by
\begin{equation}
\nabla^2 \textbf{E} + \mathrm{k}^2 \textbf{E} = 0, \quad \nabla^2 \textbf{H} + \mathrm{k}^2\textbf{H} = 0.
\label{waveEq}
\end{equation}
Here, $k^2 = \omega^2\epsilon_{\mathrm{m}}\mu_{\mathrm{m}}$ with $\omega$ as the frequency of the incident electromagnetic wave, and, $\epsilon_{\mathrm{m}}$ and $\mu_{\mathrm{m}}$ as the permittivity and the permeability of the medium, respectively. The time-harmonic electric and magnetic fields $(\textbf{E}, \textbf{H})$ associated with the incident electromagnetic wave are divergence-free in the absence of free charges or currents and related to each other through 
\begin{equation}
    \nabla \times \textbf{E} = i\omega\mu_{\mathrm{m}}\textbf{H}, \quad \nabla \times \textbf{H} = -i\omega\epsilon_{\mathrm{m}}\textbf{E}.
\end{equation} 
The scalar wave equation, resulting from Equation~(\ref{waveEq}), in cylindrical coordinates, for scattering from a long nanowire is given as
\begin{equation}
    \frac{1}{r}\frac{\partial}{\partial r}\bigg(r\frac{\partial \psi}{\partial r}\bigg) + \frac{1}{r^2}\frac{\partial^2\psi}{\partial \phi^2} + \frac{\partial^2 \psi}{\partial z^2} + k^2\psi = 0
\end{equation}
and it can be solved using separation of variables 
\begin{equation}
    \psi_{n}(r, \phi, z) = Z_{n}(\rho)e^{in\phi}e^{ihz}; \quad (n = 0, \pm 1,...).
\end{equation}
Orthogonality of vector cylindrical harmonics of integral order $n$ is ensured by
\begin{equation}
    \textbf{M}_{n} = \nabla \times (\hat{\textbf{e}}_{z}\psi_{n}), \quad \textbf{N}_{n} = \frac{\nabla \times \textbf{M}_{n}}{k}.
\end{equation}
Here, $\rho = r\sqrt{k^2 - s^2}$ and $Z_n(\rho)$ is a solution to the Bessel equation with Bessel functions of the first ($J_n$) and the second ($Y_n$) kind as the linearly independent solutions of integral order $n$. 
The separation constant $s$ is identified based on the form of the incident electromagnetic wave and the requirement of satisfying the necessary boundary conditions for it at the interface between the cylinder and the host medium that it is embedded in. 
For a long right circular cylinder of radius $a$ with a plane homogeneous wave $\textbf{E}_{i} = \textbf{E}_{0}e^{ik\hat{\textbf{e}}_{i}\cdot\textbf{x}}$ incident at an angle $\zeta$ and propagating in the direction $\hat{\textbf{e}_i} = -\sin(\zeta)\hat{\textbf{e}}_x - \cos(\zeta)\hat{\textbf{e}}_z$, we consider two polarization states of the incident plane electromagnetic wave such that its electric field is polarized \textbf{(a)} parallel (TM), and, \textbf{(b)} perpendicular (TE) to the $xz$-plane.

Briefly, employing the orthogonality property of the vector cylindrical harmonics in conjunction with the requirement for the expansion to be finite at $r = 0$, the incident electric field $\textbf{E}_{\rm{inc}}^{\rm{par}}$ parallel to the $xz$-plane and the associated magnetic field $\textbf{H}_{\rm{inc}}^{\rm{par}}$ can be expanded as
\begin{equation}
    \textbf{E}_{\rm{inc}}^{\rm{par}} = \sum_{n=-\infty}^{\infty} E_n \textbf{N}_n; \quad \textbf{H}_{\rm{inc}}^{\rm{par}} = \frac{-ik}{\omega \mu}\sum_{n=-\infty}^{\infty} E_n \textbf{M}_n,
\end{equation}
where $E_n = E_{\rm{o}}(-i)^n/k \sin{\zeta}$ and the vector harmonics $\textbf{N}_n$ and $\textbf{M}_n$ are functions of the cylindrical Bessel function $J_n(kr\sin{\zeta})e^{in\phi}e^{-ikz\cos{\zeta}}$. Similar to the expansion of the incident electromagnetic fields, the internal fields ($\textbf{E}_{\rm{int}}^{\rm{par}}$, $\textbf{H}_{\rm{int}}^{\rm{par}}$) can be expanded using the generating functions $J_n(kr\sqrt{m^2 - \cos^2{\zeta}})e^{in\phi}e^{-ikz\cos{\zeta}}$ as
\begin{equation}
    \textbf{E}_{\rm{int}}^{\rm{par}} = \sum_{n=-\infty}^{\infty} E_n[g_{n}^{\rm{par}}\textbf{M}_n + f_{n}^{\rm{par}}\textbf{N}_n]; \quad
    \textbf{H}_{\rm{int}} = \frac{-ik_{\rm{int}}}{\omega \mu_{\rm{int}}}\sum_{n=-\infty}^{\infty}E_n[g_{n}^{\rm{par}}\textbf{N}_n + f_{n}^{\rm{par}}\textbf{M}_n], 
\end{equation}
where $m$, $\mu_{\rm{int}}$ and $k_{\rm{int}}$ are the relative refractive index of the cylinder with respect to the host medium, its magnetic permeability, and the wavevector of the electromagnetic field in the cylindrical wire, respectively. Here again, the use of the cylindrical Bessel functions as the generators of the solution for the internal fields in the long nanowire guarantees that the solution is well-behaved at the origin. Next, to describe the scattered fields ($\textbf{E}_{\rm{sca}}^{\rm{par}}$, $\textbf{H}_{\rm{sca}}^{\rm{par}}$) that must constitute an outgoing wave at large distances from the scatterer, Hankel functions of the first kind ($H_n = J_n + iY_n = H_n(kr\sin{\zeta})$) are employed, 
\begin{equation}
    \textbf{E}_{\rm{sca}}^{\rm{par}} = -\sum_{n=-\infty}^{\infty} E_n[b_{n}^{\rm{par}}\textbf{N}_n + i a_{n}^{\rm{par}}\textbf{M}_n]; \quad
    \textbf{H}_{\rm{sca}}^{\rm{par}} = \frac{ik}{\omega \mu}\sum_{n=-\infty}^{\infty}E_n[i b_{n}^{\rm{par}}\textbf{M}_n + i a_{n}^{\rm{par}}\textbf{N}_n].
\end{equation}
Asymptotically, for $|\rho| \gg n^2$ Hankel functions of the first kind become $H_n(\rho) = (\sqrt{2/\pi\rho})e^{i\rho}(-1)^ne^{-i\pi/4}$ thereby ensuring an outgoing scattered wave. The coefficients of expansion for the internal ($\textbf{E}_{\mathrm{int}}$, $\textbf{H}_{\mathrm{int}}$) and the scattered fields ($\textbf{E}_{\mathrm{sca}}$, $\textbf{H}_{\mathrm{sca}}$) are obtained using the fields ($\textbf{E}_{\mathrm{inc}}$, $\textbf{H}_{\mathrm{inc}}$) of the incident electromagnetic radiation by ensuring a continuity of the tangential components of the fields upon an application of the boundary conditions 
\begin{equation}
(\textbf{E}_{\mathrm{inc}} + \textbf{E}_{\mathrm{sca}} - \textbf{E}_{\mathrm{int}}) \times \hat{\textbf{e}}_{r} = (\textbf{H}_{\mathrm{inc}} + \textbf{H}_{\mathrm{sca}} - \textbf{H}_{\mathrm{int}}) \times \hat{\textbf{e}}_{r} = 0
\label{BCs}
\end{equation}
at the interface ($r = a$) between the cylinder and the surrounding medium. The resulting four equations for the expansion coefficients can be solved for the radiation incident normal to the cylinder axis ($\zeta = 90 \degree$) giving
\begin{equation}
    a_{-n}^{\rm{par}} = a_{n}^{\rm par} = 0; \quad b_{-n}^{\rm par} = b_{n}^{\rm par} = \frac{J_n(mx)J_{n}^{'}(x) - mJ_{n}^{'}(mx)J_{n}(x)}{J_n(mx)H_{n}^{'}(x) - mJ_{n}^{'}(mx)H_{n}(x)}.
\label{abPar}    
\end{equation}
Here, the cylinder and the host materials are assumed to be non-magnetic with relative magnetic permeabilities ($\mu = \mu_{\rm{host}})$ taken as unity and $x = ka$. Similarly, the electric field $\textbf{E}_i = E_0\hat{\textbf{e}}_y e^{-ik(r\sin{\zeta}\cos{\phi} + z\cos{\zeta})}$ of the normally ($\zeta = 90 \degree$) incident radiation perpendicular to the $xz$-plane can be expanded as 
\begin{equation}
    \textbf{E}_{\rm{inc}} = -i\sum_{n=-\infty}^{\infty} E_n \textbf{M}_n,
\end{equation}
and the coefficients of expansion ($a_{n}^{\rm{per}}, b_{n}^{\rm{per}}$) for the scattered electric field
\begin{equation}
    \textbf{E}_{\rm{sca}} = \sum_{n=-\infty}^{\infty} E_n(ia_{n}^{\rm{per}} \textbf{M}_n + b_{n}^{\rm{per}} \textbf{N}_n)
\end{equation}
can be shown to be 
\begin{equation}
    b_{-n}^{\rm{per}} = b_{n}^{\rm{per}} = 0; \quad a_{-n}^{\rm{per}} = a_{n}^{\rm{per}} = \frac{mJ_{n}^{'}(x)J_{n}(mx) - J_{n}(x)J_{n}^{'}(mx)}{mJ_{n}(mx)H_{n}^{'}(x) - J_{n}^{'}(mx)H_{n}(x)}.
\label{abPer}
\end{equation}
Now, employing the scattering and absorption cross-sections per unit length of the long cylinder, the corresponding efficiencies for the parallel (TM) and perpendicular (TE) polarizations of the incident radiation can be written as
\begin{equation}
    Q_{\rm{sca}}^{\rm{par}} = \frac{2}{x}\bigg{[} {|b_{0}^{\rm{par}}|}^2 + 2\sum_{n=1}^{\infty} {|b_{n}^{\rm{par}}|}^2 \bigg{]}; 
\label{Qscapar}
\end{equation}
\begin{equation}
    Q_{\rm{abs}}^{\rm{par}} = \frac{2}{x} \bigg{[} {\rm{Re}} \bigg{(} b_{0}^{\rm{par}} + 2\sum_{n=1}^{\infty} b_{n}^{\rm{par}} \bigg{)}  - \bigg{(} {|b_{0}^{\rm{par}}|}^2 + 2\sum_{n=1}^{\infty} {|b_{n}^{\rm{par}}|}^2\bigg{)}  \bigg{]};
\label{Qabspar}    
\end{equation}
\begin{equation}
    Q_{\rm{sca}}^{\rm{per}} = \frac{2}{x}\bigg{[} {|a_{0}^{\rm{per}}|}^2 + 2\sum_{n=1}^{\infty} {|a_{n}^{\rm{per}}|}^2 \bigg{]};
\label{Qscaper}
\end{equation}
\begin{equation}
    Q_{\rm{abs}}^{\rm{per}} = \frac{2}{x} \bigg{[} {\rm{Re}} \bigg{(} a_{0}^{\rm{per}} + 2\sum_{n=1}^{\infty} a_{n}^{\rm{per}} \bigg{)}  - \bigg{(} {|a_{0}^{\rm{per}}|}^2 + 2\sum_{n=1}^{\infty} {|a_{n}^{\rm{per}}|}^2\bigg{)} \bigg{]}.
\label{Qabsper}
\end{equation}
For the computation of the above scattering and absorption efficiencies, we make use of an algorithm that relies on the logarithmic derivative $D_n(\rho)= J_{n}^{'}(\rho)/J_n(\rho)$ and the recurrence relation for the Bessel functions $Z_{n}^{'}(x) = Z_{n-1}(x) - (n/x)Z_n(x)$ to calculate the expansion coefficients $b_{n}^\mathrm{par}$ and $a_{n}^\mathrm{per}$. The summation of the terms in Equations~(\ref{Qscapar})-(\ref{Qabsper}) is truncated after $n_{\rm{stop}} \approx x + 4x^{1/3} + 2$ terms. For more details on the derivation of the results outlined above and the algorithm itself the reader is referred to the text by Bohren and Huffman \cite{RN102}.

\subsection*{Complex refractive indices}
For the calculation of the temperature-dependent dielectric permittivities of the three materials, we employ models proposed by Rakic et al., Green et al. and Reinhart for Au, Si and GaAs, respectively, as reported in our previous publication  \cite{RN159}. We also explicitly account for the radial expansion of the nanowires at elevated temperatures \cite{RN159}. However, it is assumed that the nanowires are in thermal equilibrium with the surrounding host material at any given temperature and thermal transients are absent. The computation of the temperature-dependent dielectric permittivities for Au, Si and GaAs is briefly described below while the reader interested in more details is referred to the studies published earlier \cite{RN159, RN118, RN117, RN115}. 

\subsubsection*{Gold}
More specifically, we employ the Drude-Lorentz (DL) model to compute the dielectric permittivity for Au  in the wavelength range $\lambda = 400 -1450$ nm based on the model estimated by Rakic \textit{et al.} and given by \cite{RN118}
\begin{equation}
\epsilon(\omega) = \epsilon_{\infty} - \frac{\Omega_{\mathrm{p}}^{2}}{\omega^{2}+i\Gamma_{\mathrm{D}}\omega} + \sum^5_{j=1} \frac{C_j\omega_{\mathrm{p}}^{2}}{(\omega_j^2 - \omega^2) + i\omega\gamma_j}.
\label{epsilonDL}
\end{equation}
Here, $\Omega_\mathrm{p} = \sqrt{f_\mathrm{0}}\omega_\mathrm{p}$ is the plasma frequency for the intraband transitions of oscillator strength $f_\mathrm{0}$ and the Drude damping factor $\Gamma_\mathrm{D}$; $\epsilon_{\infty}$, $\omega$ and $\omega_{\mathrm{p}}$ are the background dielectric constant, frequency of the incident radiation and the plasma frequency, respectively. The parameters $C_j$, $\omega_j$, and $\gamma_j$ represent the oscillator strength, characteristic frequency and damping respectively of the Lorentz oscillators included in the DL model \cite{RN118}. The second term in Equation~\eqref{epsilonDL} models the temperature dependence of the permittivity through the plasma frequency $\Omega_\mathrm{p}$ and the Drude damping factor $\Gamma_\mathrm{D}$. The Lorentz oscillator parameters $C_j$, $\omega_j$ and $\gamma_j$ in the DL model are estimated from ellipsometric measurements and are assumed to have no temperature dependence \cite{RN118}. The temperature dependence of the plasma frequency $\omega_\mathrm{p}$ [$=Ne^2/m^*\epsilon_\mathrm{0}$] relates to a change in the carrier concentration $N$ due to lattice thermal expansion and is given by \cite{RN147}
\begin{equation}
\omega_\mathrm{p}(T) = \frac{\omega_\mathrm{p}(T_\mathrm{0})}{\sqrt{1 + 3\alpha_\mathcal{L}(T - T_\mathrm{0})}},
\label{omegaT}   
\end{equation}
wherein $\alpha_\mathcal{L}$ is the coefficient of linear thermal expansion for Au, $T$ is the temperature, $T_\mathrm{0} = 293.15$ K is the reference temperature, $m^*$ is the effective mass of the electrons, $e$ is the electric charge and $\epsilon_\mathrm{0}$ is the permittivity of the free space. The Drude damping $\Gamma_\mathrm{D}$ ($= \Gamma_\mathrm{ee} + \Gamma_\mathrm{e\phi}$) has temperature dependent contributions from both electron-electron (${e-e}$) and electron-phonon (${e- \phi}$) interactions through the corresponding damping factors $\Gamma_\mathrm{ee}$ and $\Gamma_\mathrm{e\phi}$ \cite{RN147, RN132, RN119}. 

\subsubsection*{Silicon}
The temperature dependent refractive index $n_\mathrm{p}$ for Si is calculated using an empirical power law   
\begin{equation}
\zeta(T) = \zeta(T_\mathrm{0})(T/T_\mathrm{0})^{b}
\end{equation}
that is employed to model the dielectric dispersion in intrinsic Si \cite{RN117}. Here, $\zeta$ represents the real ($\eta$) or the imaginary ($\kappa$) part of the refractive index $n_\mathrm{p}$, $T$ and $T_\mathrm{0} = 300$ K are the actual and the reference temperatures, respectively, and the exponent $b$ depends on the normalized temperature coefficients $C_{\zeta}$ [$\equiv (1/\zeta)d\zeta/dT$] of the optical constants $\eta$ and $\kappa$ through
\begin{equation}
b = C_{\zeta}(T)T = C_{\zeta}(T_\mathrm{0})T_\mathrm{0}.
\end{equation}
The optical constants $\eta$ and $\kappa$ for Si are determined using accurate spectroscopic ellipsometry measurements reported by Herzinger \textit{et al.} \cite{RN148} and Green \cite{RN117}. Kramers-Kronig analysis \cite{RN152} is then employed to calculate optical constants using measured reflectance $\mathcal{R}$ to first compute its phase $\theta$ in radians at the target energy $E_\mathrm{t}$ as 
\begin{equation}
\theta(E_{\mathrm{t}}) = -\frac{E_{\mathrm{t}}}{\pi}\bigg[\int_{0}^{E_{\mathrm{t}} - \delta}\frac{\ln[\mathcal{R}(E)]}{E^2 - E_\mathrm{t}^{2}}dE + \int_{E_{\mathrm{t}} - \delta}^{\infty}\frac{\ln[\mathcal{R}(E)]}{E^2 - E_\mathrm{t}^{2}}dE\bigg],
\end{equation}
where $\delta\rightarrow 0$ and $E$ is a dimensionless quantity. The refractive index ($\eta + i\kappa$) is calculated as 
\begin{equation}
\eta(E) = \frac{1 - \mathcal{R}(E)}{1 + \mathcal{R}(E) - \sqrt{2\mathcal{R}(E)}\cos\theta};
\end{equation}
\begin{equation}
\kappa(E) = \frac{2\sqrt{\mathcal{R}(E)}\sin\theta}{1 + \mathcal{R}(E) - \sqrt{2\mathcal{R}(E)}\cos\theta},
\end{equation} 
to obtain the temperature dependent dielectric permittivity for Si given by 
\begin{equation}
\epsilon(T) = n_p^2(T) = [\eta(T) + i\kappa(T)]^2 = \eta^2 - \kappa^2 + 2i\eta\kappa.
\label{epsilonNK}
\end{equation}
\subsubsection*{GaAs}
For direct bandgap intrinsic GaAs, we use a heuristic model to account for the existence of the experimentally observed band tails and the non-parabolic shape of absorption above the bandgap by estimating the continuum ($\alpha_\mathrm{c}$) and the exciton ($\alpha_\mathrm{ex}$) contributions to the absorption coefficient $\alpha =\alpha_\mathrm{c} + \alpha_\mathrm{ex}$. This involves fitting the following functions to the experimental absorption spectrum \cite{RN115} 
\begin{equation}
\alpha_\mathrm{c}(E') = A\ \exp[r(E')]\bigg[\frac{1}{1 + \exp(-E'/E_\mathrm{s1})} + \frac{a_\mathrm{so}}{1 + \exp((\Delta-E')/E_\mathrm{so})}\bigg],
\label{alphaC}
\end{equation}
and
\begin{equation}
\alpha_\mathrm{ex}(E') = \sum_{j=1}^{2}\bigg[ \frac{a_{\mathrm{x}j}}{\exp[(E' + E_{\mathrm{x}j})/E_\mathrm{s1}] + \exp[-(E' + E_{\mathrm{x}j})/E_\mathrm{s2}]}\bigg],
\label{alphaE}
\end{equation} 
where
\begin{align*}
r(E') = r_{1}E' + r_{2}E'^{2} + r_{3}E'^{3}\ \mathrm{and}\ E' = E - E_\mathrm{g}(T).
\end{align*}
The increase in the absorption $\alpha$ above the fundamental bandgap is described by the first term in Equation~\eqref{alphaC} and the sharp step-like absorption increase at the fundamental and the split-off gaps is captured by the Fermi-function-type terms. The bandgap-shrinkage effect due to an increase in temperature is modeled using \cite{RN130}
\begin{equation}
E_\mathrm{g}(T) = E_\mathrm{g}(0) - \frac{\sigma \Theta}{2}\bigg[\sqrt[p]{1 + \bigg(\frac{2T}{\Theta}\bigg)^p}  - 1\bigg].
\end{equation}
Here, $E_\mathrm{g}(0) = 1.5192$ eV is the zero-temperature bandgap, $\sigma = 0.475$ eV/K is the high-temperature limiting value of the forbidden-gap entropy, $\Theta = 222.4$ K is the material specific phonon temperature used to represent the effective phonon energy ($\hslash\omega_\mathrm{eff}\equiv k_{\rm B}\Theta$, $\hslash$ is the reduced Planck's constant), and $p = 2.667$ is an empirical parameter that accounts for the globally concave shape ($p > 2$) of the electron-phonon spectral distribution function \cite{RN130}.

The absorption amplitudes for the fundamental and the split-off valence band transitions are given by the parameters $A$ and $a_\mathrm{so}$ in Equation~\eqref{alphaC}, respectively. The coefficients $r_j$ describe the increase in the absorption for  energies $E > E_\mathrm{g}$. The parameters $E_{\mathrm{s1}}$ and $E_\mathrm{s2}$ in Equation~\eqref{alphaE} represent the energy slope, and alongside $E_\mathrm{so}$ in Equation~\eqref{alphaC} for the split-off valence band, account for the weak electron-phonon coupling through a linear dependence on the temperature.
The discrete transition amplitudes $a_{\mathrm{x}j} \ (j = 1,2)$ in Equation~\eqref{alphaE} depend on the temperature and relate to the transitions of the excitonic ground and first excited states. 
The Kramers-Kronig integral can again be used to compute the imaginary part $\kappa$ of the refractive index $n$ from the absorption coefficient $\alpha$ as a function of the energy ($\mathcal{E}$) using \cite{RN115, RN152} 
\begin{equation}
\kappa(E) = \frac{c \hslash}{q\pi} \int_{0}^{E_\mathrm{u}}\frac{\alpha(\mathcal{E})d\mathcal{E}}{\mathcal{E}^2 - E^2}.
\end{equation}
Here, $c$ is the speed of light and $q$ is the magnitude of the electronic charge. The upper limit of the integral is specified as $E_\mathrm{u} = E_\mathrm{g} + 1.5$ eV.
The real part of the refractive index $\eta$ has the dominant contributions from the high critical points $E_2 = 3$ and $E_3 = 5$ eV and is estimated using  \cite{RN115}
\begin{equation}
\eta(E) = \sqrt{1 + \frac{A_{\mathrm{K}2}}{E_2^2 - E^2} + \frac{A_{\mathrm{K}3}}{E_3^2 - E^2} + \frac{A_4}{E_4^2 - E^2}}.
\end{equation}
The oscillator strengths ($A_{\mathrm{K}i}$ in $\mathrm{eV^2}$) for the critical points are quadratic functions of the temperature $T$ and the last term approximates the contribution of the reststrahl band with $A_4 = 0.0002382$ ($\mathrm{eV^2}$) and $E_4 = E_\mathrm{ph} = 0.0336$ eV the optical phonon energy \cite{RN115}. The parameters for the real part of the refractive index are estimated from a fitting to the precise refractive index data obtained from experiments and the dielectric permittivity for GaAs can thus be calculated using Equation~\eqref{epsilonNK}.

\section*{Results and discussion}

\subsection*{Mie efficiencies  - polarization and temperature dependence}

We first examine and compare the polarization and temperature dependence of the scattering and absorption efficiencies for the Au, Si and GaAs cylindrical nanowires as a function of their radii.

\subsubsection*{Scattering Efficiency}
\begin{figure}[ht!]	
\centering
\includegraphics[scale=0.64]{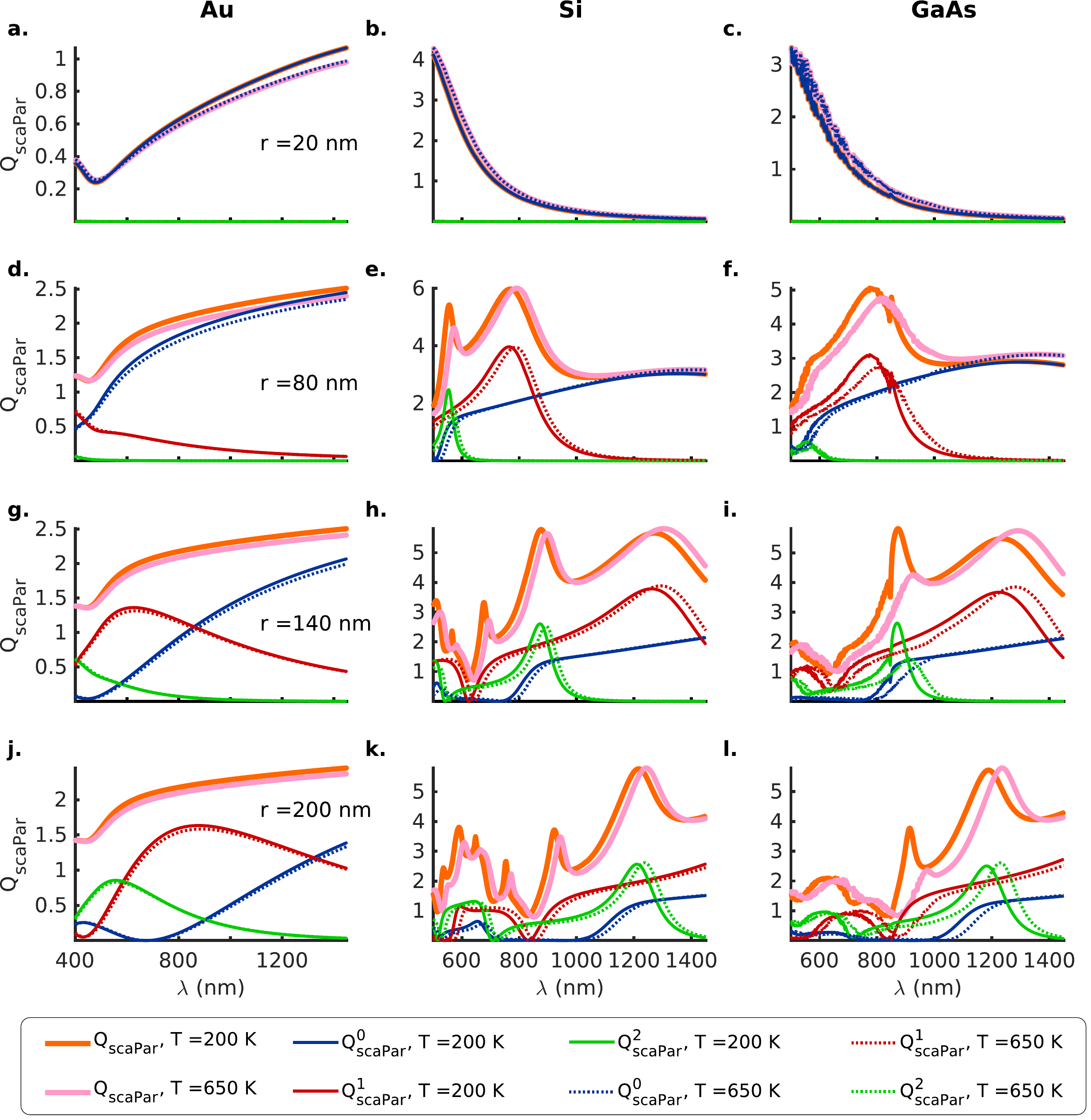} 
\caption{Mie scattering efficiency ($Q_\mathrm{scaPar}$) for the TM ($E_{\parallel}$) polarization of the incident radiation with zeroth (blue), first (red) and second (green) order contributions as a function of the wavelength, $\lambda$, for (a, d, g, j) Au (Drude-Lorentz model), (b, e, h, k) Si and (c, f, i, l) GaAs nanoparticles of radii $r = 20, 80, 140$ and $200$ nm, respectively, at temperatures $T = 200$ (solid lines) and $650$ (dotted lines) K. The thick orange and pink solid lines represent the total scattering efficiencies, $Q_\mathrm{scaPar}$, at temperatures $T = 200$ and $650$ K, respectively. Here, the Mie computations for the nanowires of different sizes take into account their thermal expansion, although the text labels indicate the values for nanowire radii at $200$ K.} 
\label{mQscaPar} 
\end{figure}

\begin{figure}[ht!]	
\centering
\includegraphics[scale=0.64]{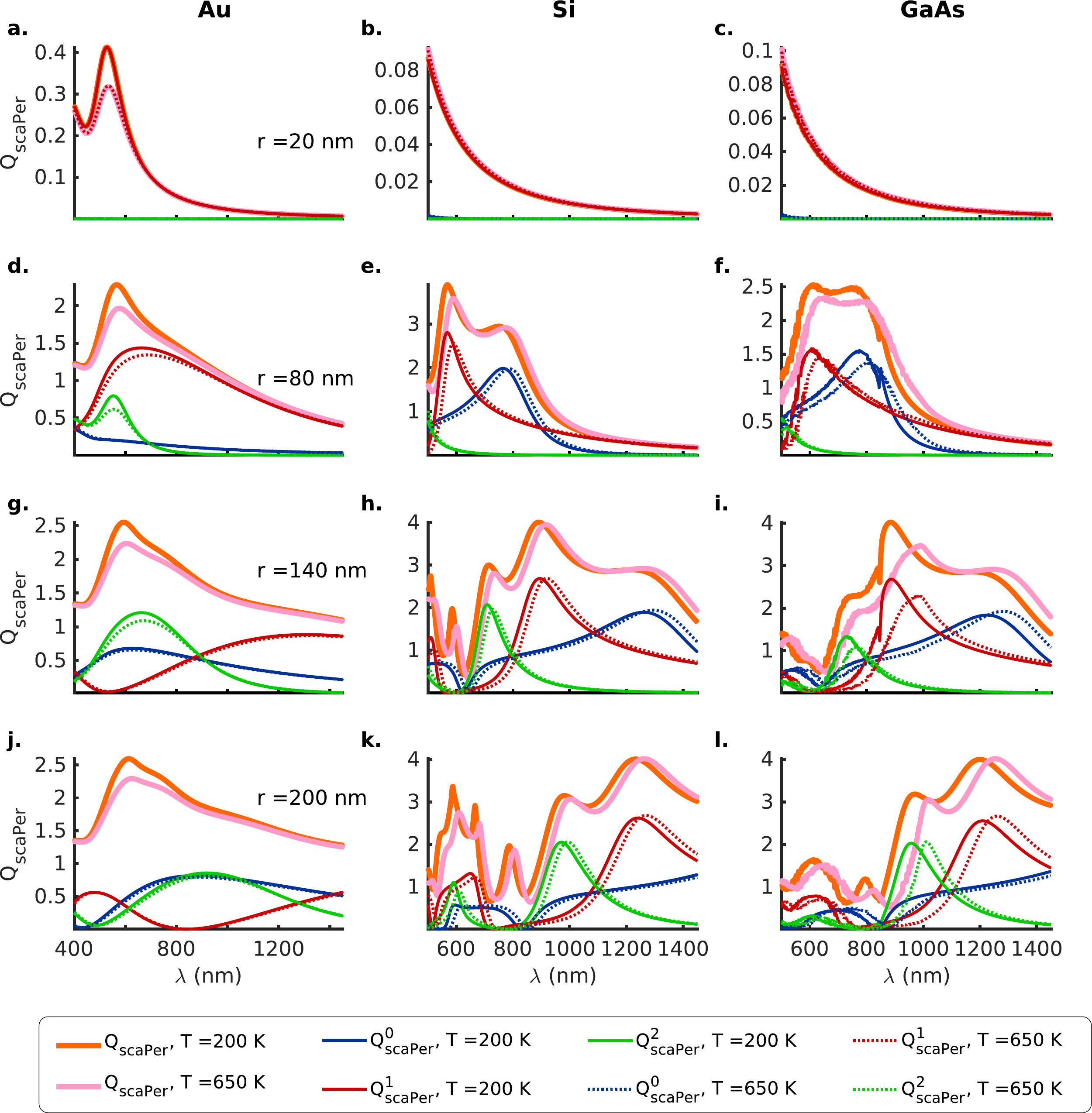} 
\caption{Mie scattering efficiency ($Q_\mathrm{scaPer}$) for the TE ($E_{\perp}$) polarization of the incident radiation with zeroth (blue), first (red) and second (green) order contributions as a function of the wavelength, $\lambda$, for (a, d, g, j) Au (DL model), (b, e, h, k) Si and (c, f, i, l) GaAs nanowires of radii $r = 20, 80, 140$ and $200$ nm, respectively, at temperatures $T = 200$ (solid lines) and $650$ (dotted lines) K. The thick orange and pink solid lines represent the total scattering efficiencies, $Q_\mathrm{scaPer}$, at temperatures $T = 200$ and $650$ K, respectively. Here, the Mie computations for the nanowires of different sizes take into account their thermal expansion, although the text labels indicate the values for nanowire radii at $200$ K.} 
\label{mQscaPer} 
\end{figure}

Figures~\ref{mQscaPar} and \ref{mQscaPer} show that the resonances in the scattering and absorption efficiencies for the TM ($E_{\parallel}$) polarization of the electric field occur at much longer wavelengths as compared to the case of the TE ($E_{\perp}$) polarization. This is a consequence of the confinement of the oscillating charge carriers along the radii of the nanowires as seen by the electric component of the incident EM wave for the TE polarization. However, the resonances are much stronger for the TM polarization as compared to those for the TE polarization of the incident EM wave. For the thinner ($\mathrm{20\,nm}$) nanowires, this difference is a couple of orders of magnitude for all materials (Figures~\ref{mQscaPar}a-c and Figures~\ref{mQscaPer}a-c). In general, the Au nanowires present very broad resonances that are weaker than the ones observed for the Si and GaAs nanowires except for the case of the TE polarization (Figure~\ref{mQscaPer}a) for the thin ($\mathrm{20\,nm}$) Au nanowire that presents a stronger resonance than the $\mathrm{20\,nm}$ semiconductor nanowires (Figures~\ref{mQscaPer}b,c). 

Fano resonances arising from all three lower order modes, against a broad background of zeroth and first-order resonances, are observed in the case of the thicker ($r \ge 80\,\mathrm{nm}$) semiconductor nanowires (Figures~\ref{mQscaPar}e, f, h, i, k, l and Figures~\ref{mQscaPer}e, f, h, i, k, l). However, there are no Fano resonances observed for the case of TM ($E_{\parallel}$) polarization in the Au nanowires in the wavelength range ($400 -1450$ nm) under consideration (Figures~\ref{mQscaPar}a, d, g, j). First and second-order resonances do occur in the wavelength range for the thicker ($140$ and $200 \mathrm{\,nm}$)  Au nanowires, but they merely serve to broaden the scattering efficiency without resulting in a decipherable resonance.  In contrast, the case of TE ($E_{\perp}$) polarization presents a broad Fano resonance arising due to the first {($ r = 80\, \mathrm{nm}$)} and second-order modes in the thicker ($r = \mathrm{140}$ and $\mathrm{200\,\mathrm{nm}}$) nanowires (Figures~\ref{mQscaPer}a, d, g, j).

In the case of the plasmonic Fano resonances for the Si and GaAs nanowires ($r \ge 80 \mathrm{nm}$) that occur at shorter wavelengths within the absorption band edge, there occurs a weakening with an increase in temperature that is larger, in general, for the second-order modes compared to the first-order resonances (Figures~\ref{mQscaPar}b, c, e, f, h, i, k, l and Figures~\ref{mQscaPer}b, c, e, f, h, i, k, l). This is also accompanied by a redshift in the resonance wavelength that results from a reduced interband energy gap for the electronic transitions at elevated temperatures. A second contribution to the redshift comes from a weakening of the Coulombic interaction between the oscillating charge carriers on account of thermal expansion of the nanowires.

On the other hand, for the dielectric Fano resonances occurring at longer wavelengths beyond the absorption band edge, there occurs a strengthening of the resonances with temperature and is readily apparent in the case of the $\mathrm{140\,nm}$ Si and GaAs nanowires for both polarizations of the incident EM radiation (Figures~\ref{mQscaPar}b, c, e, f, h, i, k, l and \ref{mQscaPer}b, c, e, f, h, i, k, l). The strengthening of the dielectric resonances with an increase in temperature can be attributed to the increase in the number density of the phonon modes at higher temperatures that are non-dissipative in nature. Again, similar to the plasmonic Fano resonances, a redshift of the resonance wavelengths for the nanowires is observed that is likely due to a weakening of the phonon modes as a result of thermal expansion from an increase in temperature.     
\subsubsection*{Absorption Efficiency}

\begin{figure}[ht!]	
\centering
\includegraphics[scale=0.64]{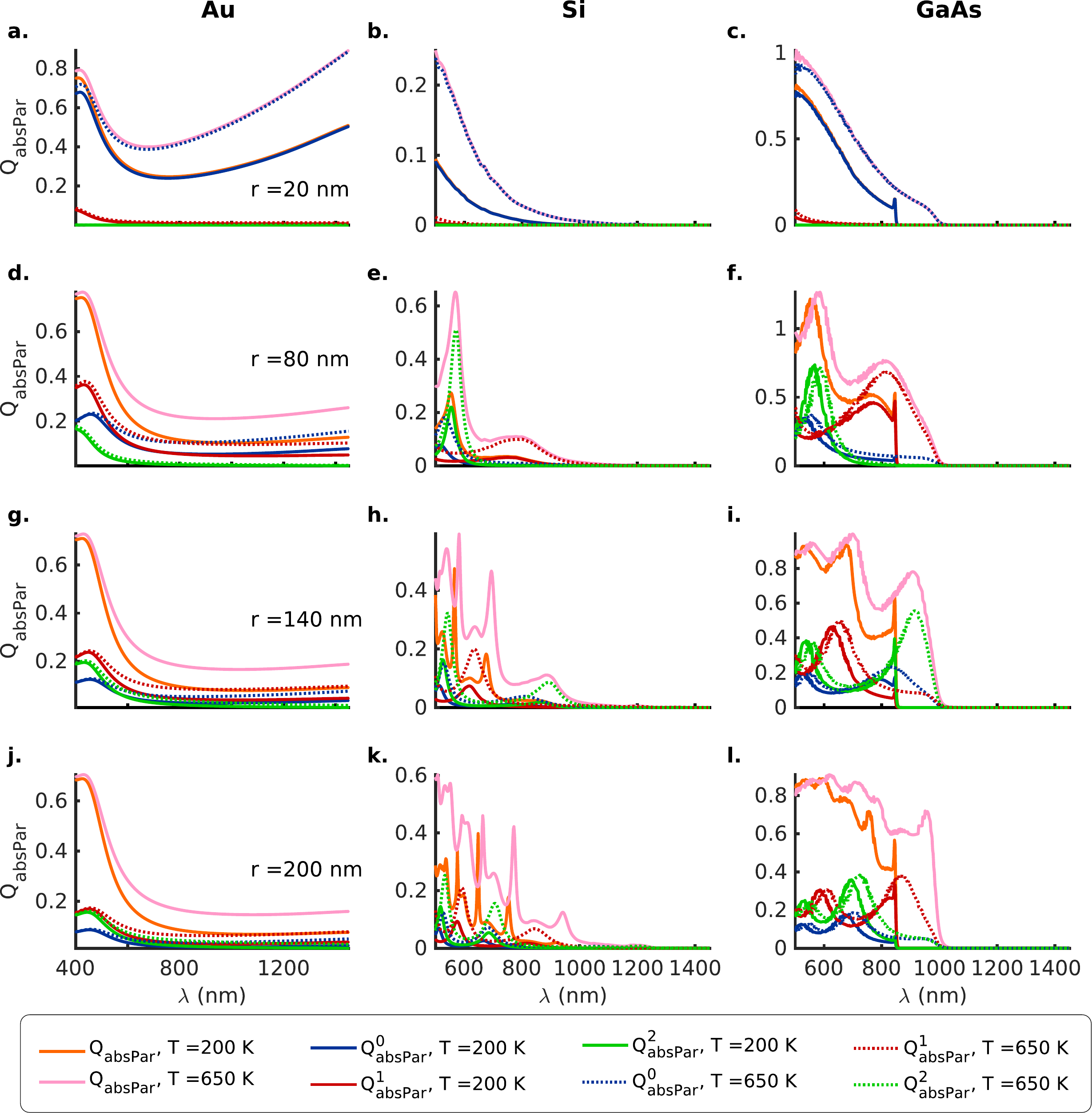} 
\caption{Mie absorption efficiency ($Q_\mathrm{absPar}$) for the TM ($E_{\parallel}$) polarization of the incident radiation with zeroth (blue), first (red) and second (green) order contributions as a function of the wavelength $\lambda$ for (a, d, g, j) Au (DL model), (b, e, h, k) Si and (c, f, i, l) GaAs nanowires of radii $r = 20, 80, 140$ and $200$ nm, respectively, at temperatures $T = 200$ (solid lines) and $650$ (dotted lines) K. The thick orange and pink solid lines represent the total absorption efficiencies, $Q_\mathrm{absPar}$, at temperatures $T = 200$ and $650$ K, respectively. Here, the Mie computations for the nanowires of different sizes take into account their thermal expansion, although the text labels indicate the values for nanowire radii at $200$ K.} 
\label{mQabsPar} 
\end{figure}

\begin{figure}[ht!]	
\centering
\includegraphics[scale=0.64]{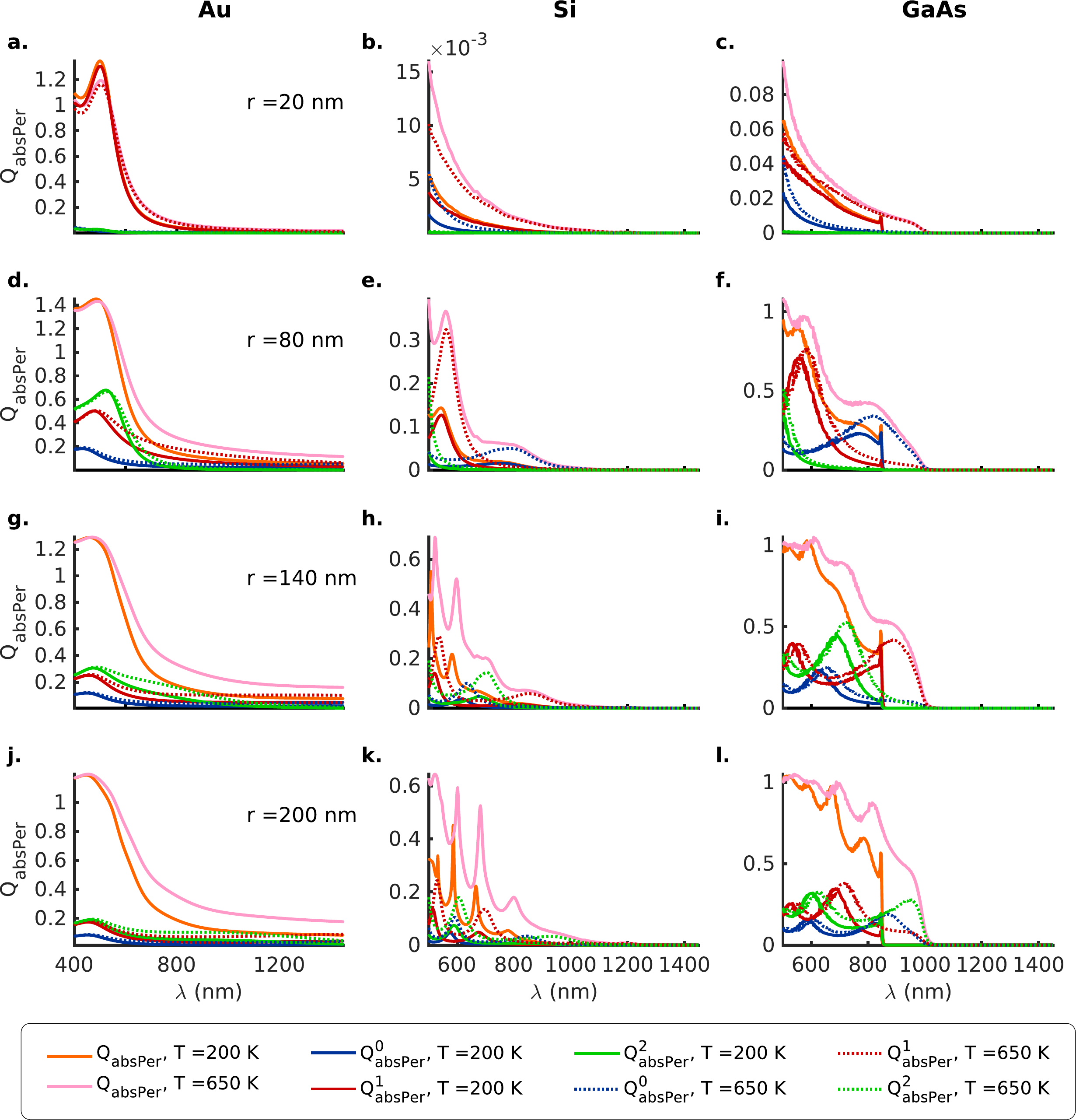} 
\caption{Mie absorption efficiency ($Q_\mathrm{absPer}$) for the TE ($E_{\perp}$) polarization of the incident radiation  with zeroth (blue), first (red) and second (green) order contributions as a function of the wavelength $\lambda$ for (a, d, g, j) Au (DL model), (b, e, h, k) Si and (c, f, i, l) GaAs nanowires of radii $r = 20, 80, 140$ and $200$ nm, respectively, at temperatures $T = 200$ (solid lines) and $650$ (dotted lines) K. The thick orange and pink solid lines represent the total absorption efficiencies, $Q_\mathrm{absPer}$, at temperatures $T = 200$ and $650$ K, respectively. Here, the Mie computations for the nanowires of different sizes take into account their thermal expansion, although the text labels indicate the values for nanowire radii at $200$ K.} 
\label{mQabsPer} 
\end{figure}

The contribution to the plasmonic absorption resonances within the absorption band at shorter wavelengths comes primarily from the zeroth-order mode for the thinnest nanowires ($\mathrm{20\, nm}$) for the three materials and the two polarizations (Figures~\ref{mQabsPar}a-c and \ref{mQabsPer}a-c). Furthermore, it is generally observed that just the three lower-order modes (zeroth, first and second) are not enough to account for all the features observed in the total absorption efficiencies ($Q_\mathrm{absPar}$, $Q_\mathrm{absPer}$). This is most evident in the thickest ($\mathrm{200\,nm}$) Si nanowire wherein the sharp resonance peaks in $Q_\mathrm{absPer}$ ($Q_\mathrm{absPar}$) at $\mathrm{666\,nm}$ ($\mathrm{754\,nm}$) and $\mathrm{682\,nm}$ ($\mathrm{772\,nm}$) for the temperatures $\mathrm{200\,}$ and $\mathrm{650\,K}$ cannot be ascribed exclusively to any of the contributions from the lower-order zeroth, first and second-order modes (Figures~\ref{mQabsPer}k and \ref{mQabsPar}k). For the most part though, the strength of the plasmonic absorption resonances is comparable in order of magnitude for both the TM ($E_{\parallel}$) and the TE ($E_{\perp}$) polarizations of the incident EM radiation, except for the case of the $\mathrm{20\, nm}$ Si and GaAs nanowires wherein the TM polarization presents resonances that are stronger by nearly two orders of magnitude (Figures~\ref{mQabsPar}b,c and \ref{mQabsPer}b,c).

Similar to what was observed for the spherical nanoparticles \cite{RN159}, (i) Au nanowires present a largely featureless broad band absorption at the longer wavelengths accompanied by broad resonances at shorter wavelengths (Figures~\ref{mQabsPar}a, d, g, j and \ref{mQabsPer}a, d, g, j). (ii) The thicker semiconductor nanowires ($r \ge 80 \mathrm{nm}$) present plasmonic absorption resonances (sharp for Si and broad for GaAs) for both polarizations within the absorption band that strengthen with an increase in the temperature from $\mathrm{200\,}$ to $\mathrm{650\,K}$ (Figures~\ref{mQabsPar}b-c, e-f, h-i, k-l and \ref{mQabsPer}b-c, e-f, h-i, k-l). (iii) A clear redshift of the absorption resonances is observed for the Si and GaAs nanowires that can be ascribed to the shrinking of the bandgap and thermal expansion at elevated temperatures (Figures~\ref{mQabsPar}b-c, e-f, h-i, k-l and \ref{mQabsPer}b-c, e-f, h-i, k-l).

Absorption resonances for the two polarizations (Figures~\ref{mQabsPar} and \ref{mQabsPer}) also show that the ordering (in terms of the resonance wavelength) and the strength of the higher first and second-order modes depends greatly on the radius of the semiconductor nanowire under consideration and can thus be considered to be highly morphology-dependent. For example, for the TM polarization and semiconductor nanowire radii $r = 140\,\mathrm{nm}$ the second-order resonances are stronger near the absorption band edge whereas for the nanowires with radii $r = 200\,\mathrm{nm}$ the first-order modes are stronger near the band edge (Figures~\ref{mQabsPar}h, i, k, l). On the other hand, for the case of the TE polarization, the converse is true (Figures~\ref{mQabsPer}h, i, k, l). In contrast, for the Au nanowires all the resonant absorption modes nearly coincide for both the polarizations (Figures~\ref{mQabsPar}a, d, g, j and \ref{mQabsPer}a, d, g, j). However, regardless of the polarization of the incident electromagnetic radiation, anomalous scattering or absorption cross-sections are observed for the semiconductor nanowires wherein the higher first and second-order modes present stronger extinction cross-sections than the lower zeroth-order modes for the thicker nanowires with $r \ge 140\, \mathrm{nm}$ in the wavelength range under consideration  (Figures~\ref{mQscaPar}, \ref{mQscaPer}, \ref{mQabsPar} and \ref{mQabsPer}b-c, e-f, h-i, k-l). 

In the case of the thin ($\mathrm{20\,nm}$) semiconductor nanowires, both Si and GaAs nanowires present low absorption efficiencies for the TE ($E_{\perp}$) polarization (Figures~\ref{mQabsPer}b,c) while for the TM ($E_{\parallel}$) polarization the Si nanowire (Figures~\ref{mQabsPar}b) presents an absorption efficiency that is a fraction of the value for the GaAs nanowire (Figure~\ref{mQabsPar}c). If one now considers the scattering efficiencies for the $\mathrm{20\,nm}$ case, the two semiconductors exhibit comparable low (high) values of the scattering efficiency for the TE (TM) polarizations of the incident electromagnetic radiation (Figures~\ref{mQscaPar} and \ref{mQscaPer}b,c). This implies that for unpolarized incident radiation the Si (indirect-bandgap) nanowires, in contrast to the GaAs (direct-bandgap) nanowires, shall function as highly efficient scatterers with a really low absorption cross-section. Unlike the case of scattering efficiencies for the TM and TE polarizations (Figures~\ref{mQscaPar} and \ref{mQscaPer}), the absorption efficiencies for the thicker nanowires $r \ge 80\, \mathrm{nm}$ are comparable in magnitude for the two polarizations for all materials considered here (Figures~\ref{mQabsPar} and \ref{mQabsPer}d-l).

\subsubsection*{Thin wire limit}
In the limit of a thin wire (size parameter, $x$ and $|mx| \ll 1$), the Equations~\eqref{abPar} and \eqref{abPer} for the Mie coefficients $b_0$ and $a_0$ corresponding to the zeroth-order modes for the TM and TE polarizations of the incident EM wave, respectively, can be approximated as \cite{RN102}
\begin{equation} 
 b_0 \simeq \frac{-i\pi x^2(m^2 - 1)}{4} \quad \mathrm{and} \quad
 a_0 \simeq \frac{-i\pi x^4(m^2 - 1)}{32}.
\end{equation}
Similarly, the first-order modes $b_1$ and $a_1$ from Equations~\eqref{abPar} and \eqref{abPer} for the TM and TE polarizations of the incident EM wave, respectively, can also be approximated as \cite{RN102}
\begin{equation} 
 b_1 \simeq \frac{-i\pi x^4(m^2 - 1)}{32} \quad \mathrm{and} \quad
 a_1 \simeq \frac{-i\pi x^2}{4}\frac{m^2 - 1}{m^2 + 1}.
\end{equation} 
As a consequence of the above relations, for the case of the thin $\mathrm{20\,nm}$ wires the zeroth-order modes $b_0$ contribute the strongest to the scattering and absorption efficiencies for the TM polarization of the incident EM radiation for all three materials (Figures~\ref{mQscaPar} and \ref{mQabsPar}a-c). In contrast, it is the first-order modes $a_1$  that make the strongest contribution to the Mie efficiencies for the TE polarization of the EM wave incident on the thin ($r = \mathrm{20\,nm}$) nanowires (Figures~\ref{mQscaPer} and \ref{mQabsPer}a-c). Furthermore, a comparison of the Figures ~\ref{mQscaPar} and \ref{mQabsPar}b (\ref{mQscaPer} and \ref{mQabsPer}b) for Si nanowires shows that the contributions of the $b_0$ mode to the scattering (absorption) efficiency for the TM polarization are roughly two orders of magnitude stronger than those from the $a_1$ mode for the TE polarization.

\subsection*{Anapole modes}
\begin{figure}[ht!]	
\centering
\includegraphics[scale=0.84]{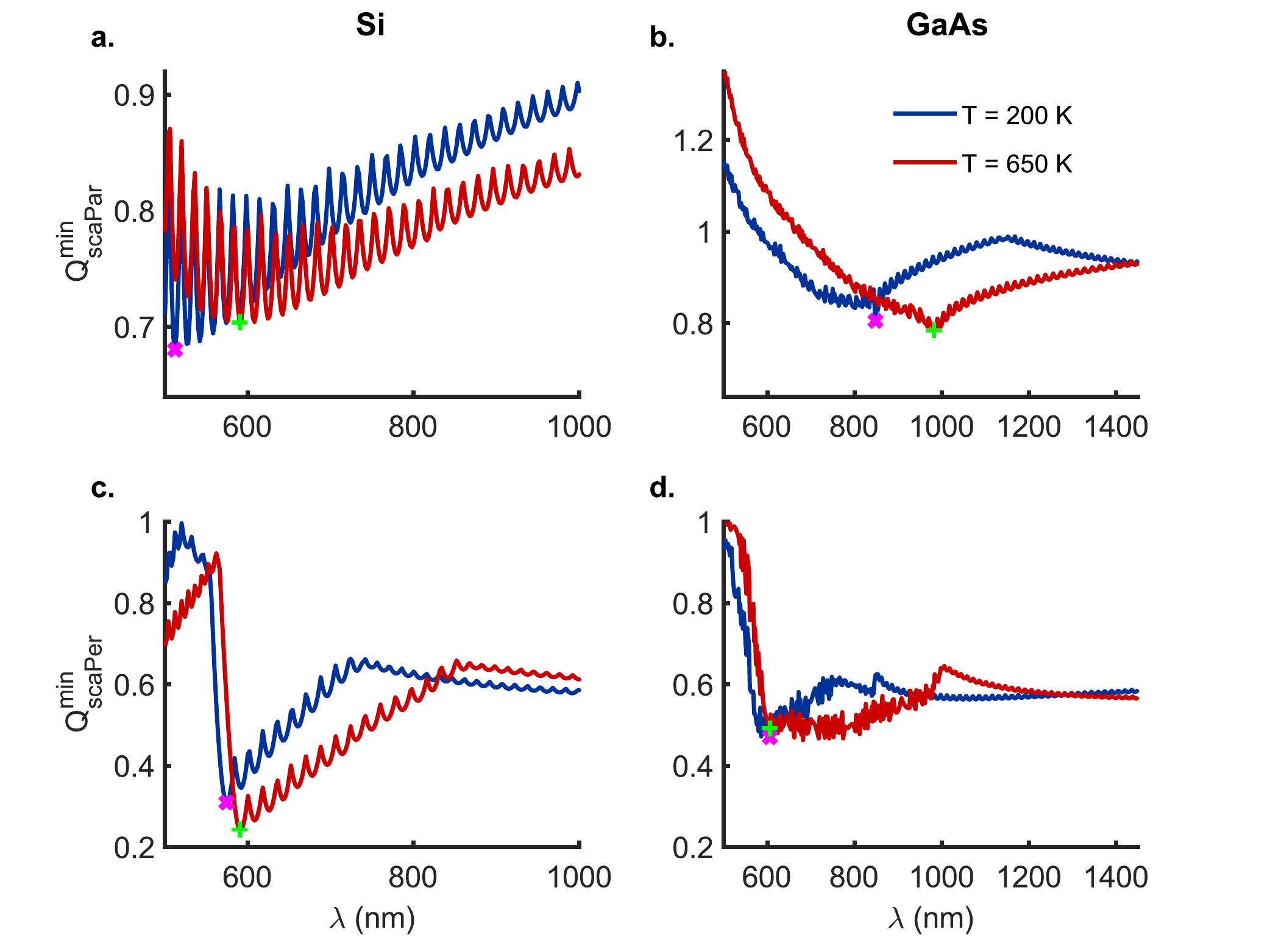} 
\caption{Spectral trajectory of $Q_\mathrm{sca}^\mathrm{min}$ for the TM ($E_{\parallel}$) (a) Si, (b) GaAs, and, TE ($E_{\perp}$) polarizations (c) Si and (d) GaAs along the minima (dashed white line) plotted in Figures S1 and S2 for $Q_{\mathrm{scaPar}}$ and $Q_{\mathrm{scaPer}}$ as a function of the nanowire radius and the wavelength of the incident radiation for the temperatures $T = 200 \,\mathrm{K}$ and $T = 650 \,\mathrm{K}$ respectively. The magenta (\textcolor{magenta}{$\times$}) and green (\textcolor{green}{$+$}) markers denote the global minima in scattering efficiencies for nanowire radii ($r > 50$ and $125 \,\mathrm{nm}$ for TM  and TE polarizations, respectively) that exhibit distinct scattering resonances as opposed to the monotonic spectral behavior observed for the smaller nanowire radii. Here, note that the points on the continuous curves shown for the two temperatures, in general, represent scattering minima at given wavelengths for nanowires with distinct radii as can be seen in the Figures~\ref{minQsca} and \ref{minQabs} wherein the radii of the nanowires corresponding to the markers (\textcolor{magenta}{$\times$}) and  (\textcolor{green}{$+$}) are clearly indicated.}  
\label{minAnapole} 
\end{figure}

Figure~\ref{minAnapole} shows the spectral trajectory of the minima in scattering efficiency $Q_\mathrm{sca}^\mathrm{min}$ for the TM (Si, Figure~\ref{minAnapole}a; GaAs, Figure~\ref{minAnapole}b), and, TE polarizations (Si, Figure~\ref{minAnapole}c; GaAs, Figure~\ref{minAnapole}d) along the minima (dashed white line) plotted in Figures~S1 (b, c, h, i) and S2 (b, c, h, i) for $Q_\mathrm{scaPar}$ and $Q_\mathrm{scaPer}$ as a function of the the wavelength of the incident radiation and nanowire radius at $T = 200$ and $650 \,\mathrm{K}$, respectively. In the wavelength range ($\lambda = 500 -1450$ nm) considered here, the global minima is observed to redshift with an increase in temperature for the TM polarization for both Si (Figure~\ref{minAnapole}a) and GaAs (Figure~\ref{minAnapole}b) nanowires. The spectral trajectory of the scattering minima exhibits an oscillatory behavior for the TM polarization in the case of Si nanowires while for the GaAs nanowires it shows a near monotonic decrease until the points of global minima at $\lambda = 848$ and $982$ $\,\mathrm{nm}$ for $T = 200$ (\textcolor{magenta}{$\times$}) and $650 \,\mathrm{K}$ (\textcolor{green}{$+$}), respectively. In contrast, for the case of TE polarization in Si nanowires ($r = 125 \,\mathrm{nm}$) (Figure~\ref{minAnapole}c), the redshift ($\Delta\lambda = 16 \,\mathrm{nm}$) of the global minima with increase in temperature from $T = 200$ (\textcolor{magenta}{$\times$}) to $650 \,\mathrm{K}$ (\textcolor{green}{$+$}) is very small with a change from $\lambda = 574$ to $590 \,\mathrm{nm}$. However, the GaAs nanowires with radii ($r = 130$ and $125 \,\mathrm{nm}$) exhibit coincident global minima in $Q_\mathrm{scaPer}$ at $\lambda = 606 \,\mathrm{nm}$ for the TE polarization. 

\begin{figure}[ht!]	
\centering
\includegraphics[scale=0.84]{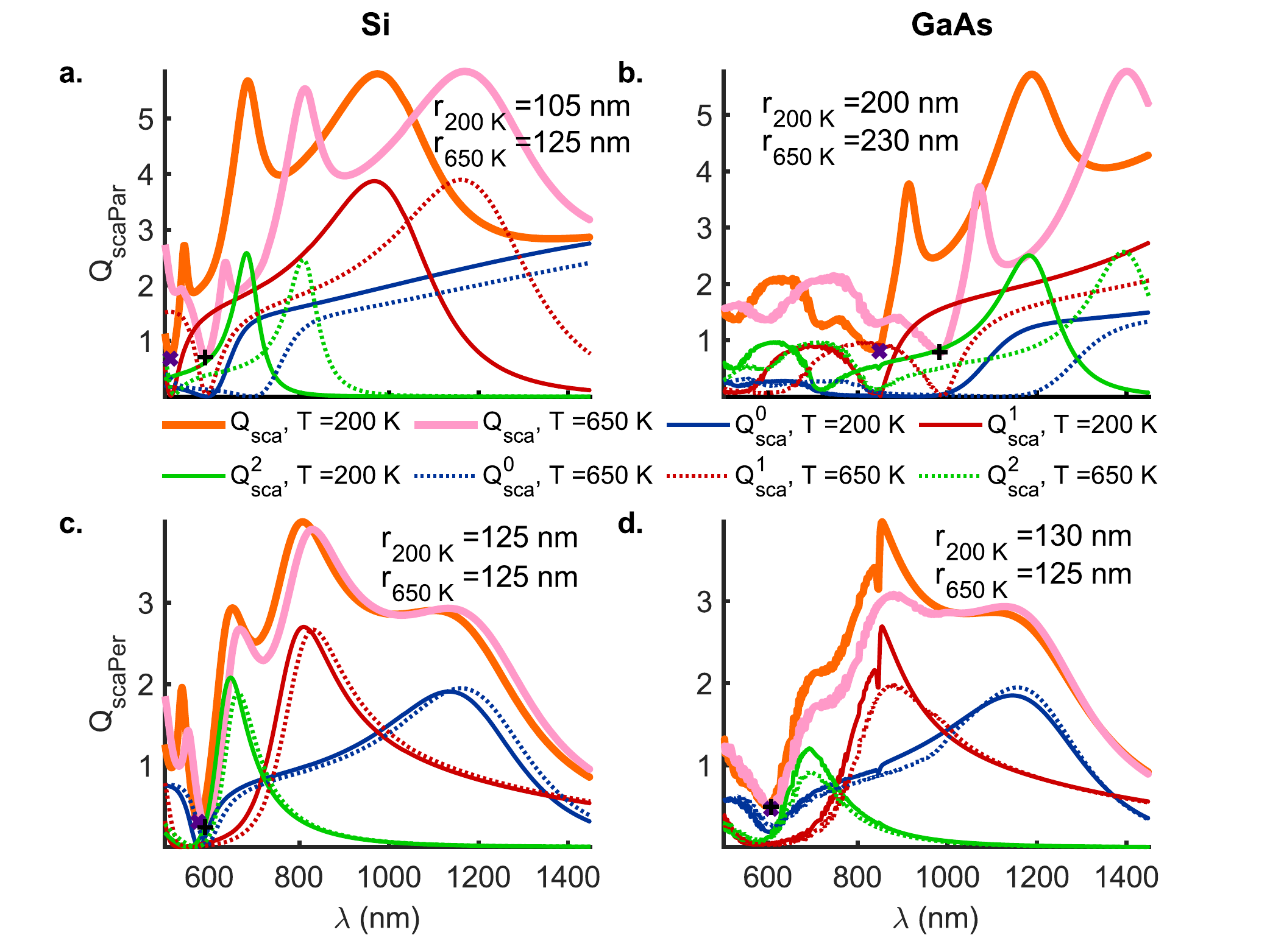} 
\caption{The scattering efficiencies $Q_\mathrm{sca}$ for the TM ($E_{\parallel}$)  (a) Si, (b) GaAs, and, TE ($E_{\perp}$) polarizations (c) Si and (d) GaAs with zeroth (blue), first (red) and second (green) order contributions as a function of the wavelength $\lambda$ at temperatures $T = 200$ (solid lines) and $650$ (dotted lines) K. The thick orange and pink solid lines represent the total scattering efficiencies, $Q_\mathrm{sca}$, at temperatures $T = 200$ and $650$ K, respectively. The radii of the nanowires considered here correspond to the global minima shown in Figure~\ref{minAnapole} for $Q_\mathrm{sca}^\mathrm{min}$ as a function of the wavelength of the incident radiation. The purple (\textcolor{violet}{$\times$}) and black (\textcolor{black}{$+$}) markers correspond to the position of the global minima in scattering efficiencies for nanowire radii ($r > 50$ and $125 \,\mathrm{nm}$) for TM and TE polarizations, respectively (See Figure~\ref{minAnapole}). Also, see Figures~S1 and S2
for a full color-map of the absoprtion efficiency $Q_\mathrm{abs}$ for the TM and TE polarizations, respectively, of the incident radiation for Au, Si and GaAs nanowires as a function of the wavelength $\lambda$ and radii $r$ at temperatures $T = 200$, $470$ and $650$ K.}
\label{minQsca} 
\end{figure}

\begin{figure}[ht!]	
\centering
\includegraphics[scale=0.84]{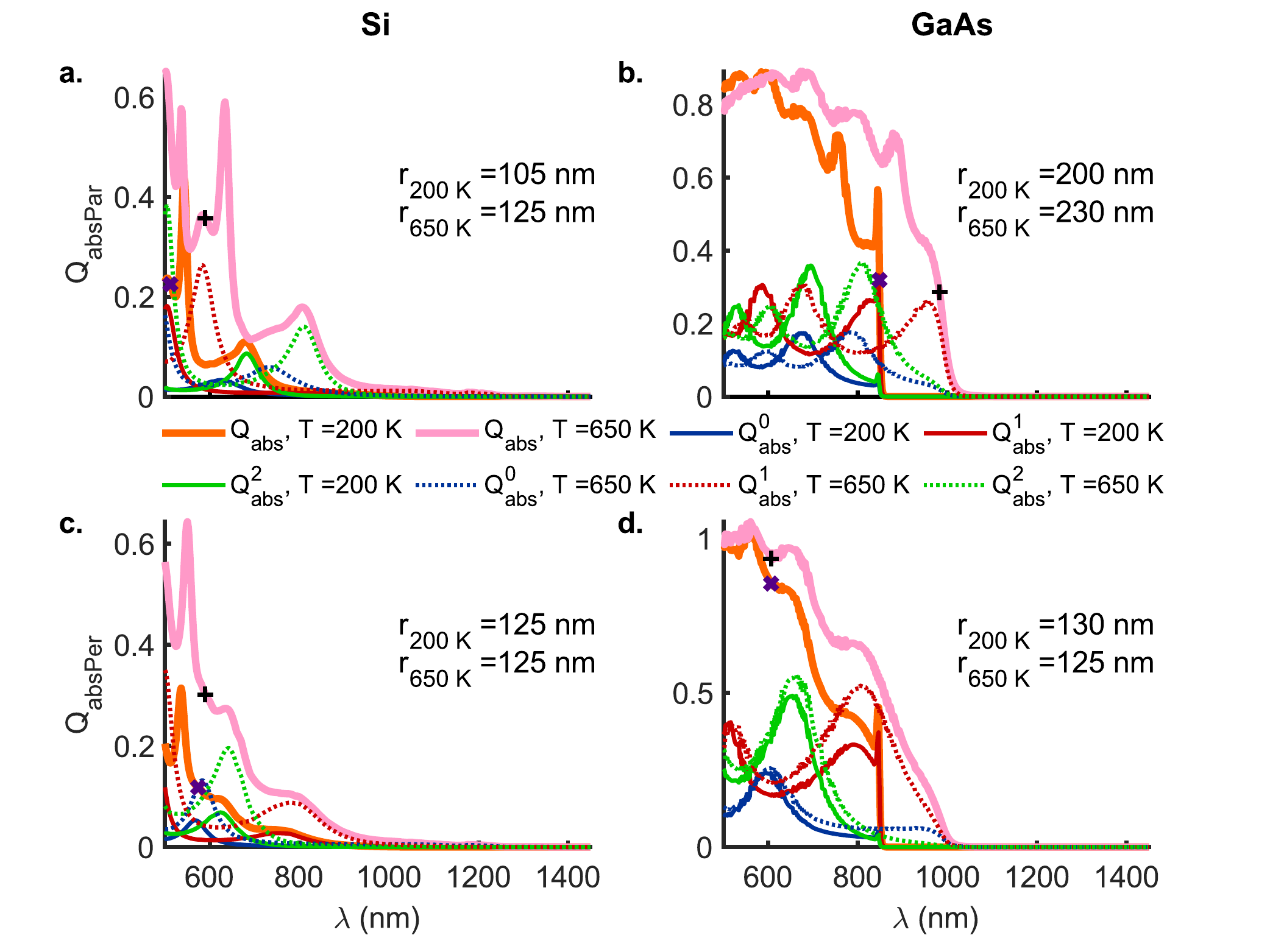} 
\caption{The absoprtion efficiencies $Q_\mathrm{abs}$ for the TM ($E_{\parallel}$) (a) Si, (b) GaAs, and, TE ($E_{\perp}$) polarizations (c) Si and (d) GaAs with zeroth (blue), first (red) and second (green) order contributions as a function of the wavelength $\lambda$ at temperatures $T = 200$ (solid lines) and $650$ (dotted lines) K. The thick orange and pink solid lines represent the total absoprtion efficiencies, $Q_\mathrm{sca}$, at temperatures $T = 200$ and $650$ K, respectively. The radii of the nanowires considered here correspond to the global minima shown in Figure~\ref{minAnapole} for $Q_\mathrm{sca}^\mathrm{min}$ as a function of the wavelength of the incident radiation. The purple (\textcolor{violet}{$\times$}) and black (\textcolor{black}{$+$}) markers correspond to the wavelength $\lambda$ of the global minima in scattering efficiencies for nanowire radii ($r > 50$ and $125 \,\mathrm{nm}$) for TM and TE polarizations, respectively (See Figure~\ref{minAnapole}). Also, see Figures~S3 and S4
for a full color-map of the absoprtion efficiency $Q_\mathrm{abs}$ for the TM and TE polarizations, respectively, of the incident radiation for Au, Si and GaAs nanowires as a function of the wavelength $\lambda$ and radii $r$ at temperatures $T = 200$, $470$ and $650$ K.}
\label{minQabs} 
\end{figure}

Figures~\ref{minQsca} and \ref{minQabs}a-b show the scattering  ($Q_{\mathrm{scaPar}}$) and absorption ($Q_{\mathrm{absPar}}$) efficiencies for the TM polarization corresponding to the marked global scattering minima for Si and GaAs nanowires at temperatures $T = 200$ and $650 \,\mathrm{K}$ in Figures~\ref{minAnapole}a-b, respectively. A weak contribution to the minima (\textcolor{violet}{$\times$}, \textcolor{black}{$+$}) in $Q_{\mathrm{scaPar}}$ (Figure~\ref{minQsca}a) for the Si nanowires ($r_{200 \,\mathrm{K}} = 105$ and $r_{650 \,\mathrm{K}} = 125 \,\mathrm{nm}$) comes from the third-order modes as the zeroth and the first-order contributions are nearly zero. However, in the case of the GaAs nanowires ($r_{200 \,\mathrm{K}} = 200$ and $r_{650 \,\mathrm{K}} = 230 \,\mathrm{nm}$), the stronger of the weak contributions at the minima (\textcolor{violet}{$\times$}, \textcolor{black}{$+$}) for $Q_{\mathrm{scaPar}}$ (Figure~\ref{minQsca}b) comes from the first-order mode at $T = 200 \,\mathrm{K}$ while at $T = 650 \,\mathrm{K}$ the second-order mode contributes the most. On the other hand, Figure~\ref{minQabs}a shows that the primary contributor to the absorption efficiency ($Q_{\mathrm{absPar}}$) for the TM polarization in the Si nanowires is a second-order resonance that strengthens and redshifts with an increase in temperature to $650 \,\mathrm{K}$ thereby indicating a strong electron-phonon coupling, characteristic of indirect bandgap semiconductors. Similar to the Si nanowires, the dominant contribution to the absorption efficiency ($Q_{\mathrm{absPar}}$) at the point of the global scattering minima (TM-polarization) for the GaAs nanowires again comes from the second-order resonant modes (Figures~\ref{minQabs}a,b). However, unlike Si nanowires, these resonant modes (redshifted for higher temperatures) are located right at the absorption band edge of the direct-bandgap GaAs nanowires.

Figures~\ref{minQsca} and \ref{minQabs}c-d show the scattering  ($Q_{\mathrm{scaPer}}$) and absorption ($Q_{\mathrm{absPer}}$) efficiencies for the TE polarization corresponding to the marked global scattering minima for Si and GaAs nanowires at temperatures $T = 200$ and $650 \,\mathrm{K}$ in Figures~\ref{minAnapole}c-d, respectively. Figure~\ref{minQsca}c shows that for Si nanowires of radii $r = 125 \,\mathrm{nm}$ the contributions from the all the lower-order ($Q_{\mathrm{scaPer}}^{0\mathrm{-}2}$) modes approach zero at $\lambda_{200 \,\mathrm{K}} = 574$ and $\lambda_{650 \,\mathrm{K}} = 590 \,\mathrm{nm}$. The scattering efficiency at the points of minima is observed to decrease from $Q_{\mathrm{scaPer}} = 0.3096$ to $0.2429$ upon an increase in the temperature from $T = 200$ to $650 \,\mathrm{K}$. On the other hand, Figure~\ref{minQabs}c shows that the absorption efficiency ($Q_{\mathrm{absPer}}$) at the points (\textcolor{violet}{$\times$}, \textcolor{black}{$+$}) of scattering minima is quite high with the strongest contribution coming from the zeroth-order resonances. In fact, the absorption efficiency more than doubles from $Q_{\mathrm{absPer}} = 0.1179$ to 0.3015 with an increase in the temperature from $200$ to $650 \,\mathrm{K}$. The GaAs nanowires of radii $r_{200 \,\mathrm{K}} = 130$ and $r_{650 \,\mathrm{K}} = 125 \,\mathrm{nm}$ (Figure~\ref{minQsca}d) exhibit coincident scattering minima at $\lambda = 606 \,\mathrm{nm}$ with $Q_{\mathrm{scaPer}} = 0.4706$ and $0.4923$, respectively. Corresponding to these minima at $\lambda = 606 \,\mathrm{nm}$, Figure~\ref{minQabs}d shows high absorption efficiencies of $Q_{\mathrm{absPer}} = 0.8539$ and $0.9350$ at $200$ and $650 \,\mathrm{K}$, respectively, with resonant contribution from the zeroth-order modes. The occurrence of such low scattering coupled with high absorption efficiencies at the same wavelengths for the TE polarization clearly points to the existence of anapole modes in Si and GaAs nanowires that are resilient to a change in temperature over a broad range from $200$ to $650 \,\mathrm{K}$.

\subsection*{Radiative vs. dissipative damping - polarization and temperature dependence}
\subsubsection*{Scattering resonances}

\begin{figure}[ht!]	
\centering
\includegraphics[scale=0.55]{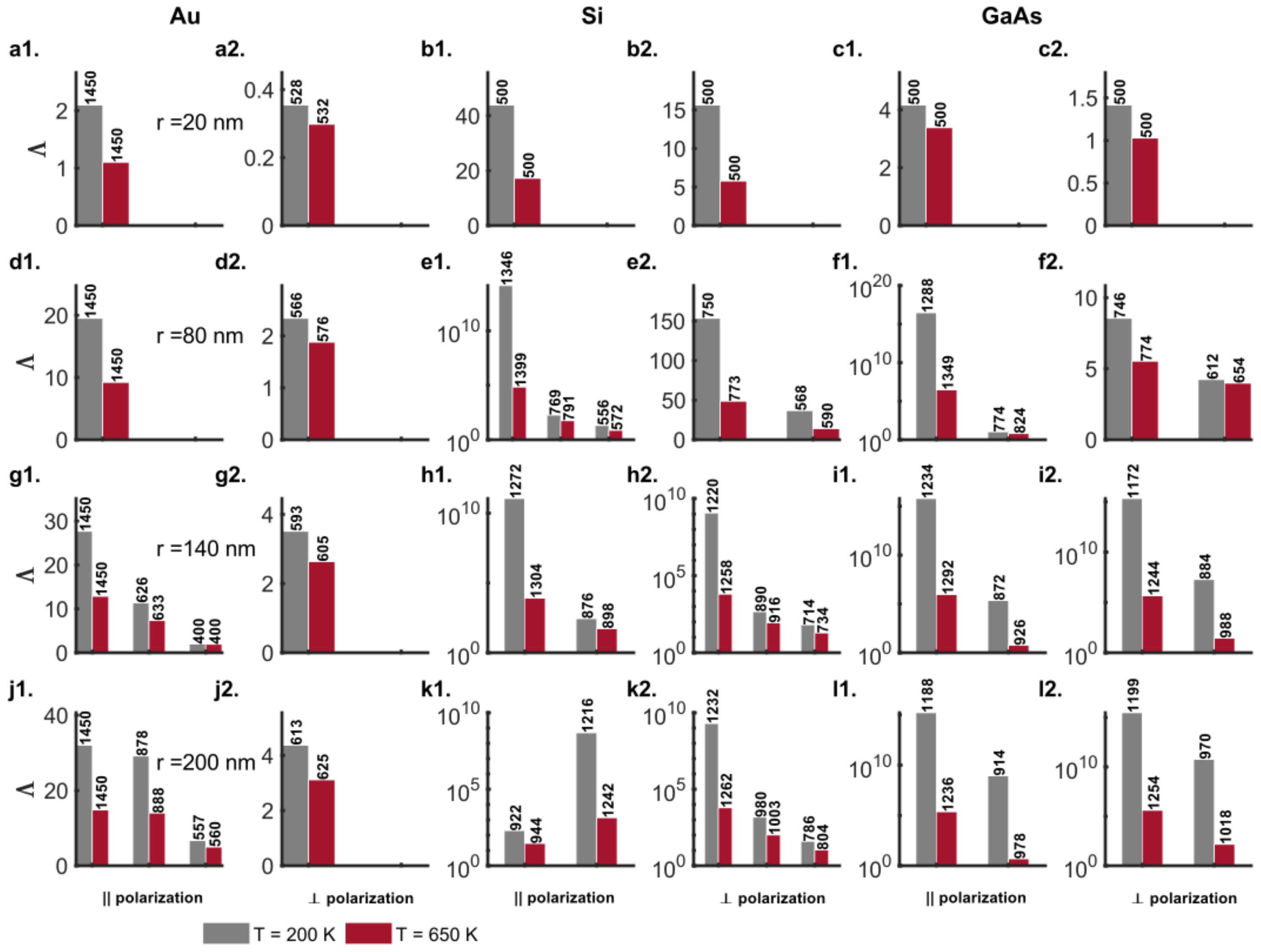} 
\caption{Ratio ($\Lambda$) of scattering efficiency ($Q_\mathrm{sca}(\lambda_{\mathrm{res}}^{\mathrm{sca}})$) to the absorption efficiency ($Q_\mathrm{abs}(\lambda_{\mathrm{res}}^{\mathrm{sca}})$) as a measure of the radiative to dissipative damping for the TM ($E_{\parallel}$) and TE ($E_{\perp}$) polarizations of the incident radiation for (a1, d1, g1, j1; a2, d2, g2, j2) Au (DL model), (b1, e1, h1, k1; b2, e2, h2, k2;) Si and (c1, f1, i1, l1; c2, f2, i2, l2) GaAs nanowires of radii $r = 20, 80, 140$ and $200$ nm, respectively, at temperatures $T = 200$ (grey bars) and $650$ (maroon bars) K. The numbers on top of the bars indicate either the wavelengths (in $\mathrm{nm}$) at which the scattering resonances occur or the maxima in scattering efficiency in the wavelength range under consideration in the case of the thin ($20 \,\mathrm{nm}$) wires. Here, the Mie computations for the nanowires of different sizes take into account their thermal expansion, although the text labels indicate the values for nanowire radii at $200$ K.} 
\label{rdScat} 
\end{figure}

Figure~\ref{rdScat} shows the ratio ($\Lambda$) of resonant scattering efficiency ($Q_\mathrm{sca}(\lambda{\mathrm{_{res}^{sca}}})$) to the absorption efficiency ($Q_\mathrm{abs}(\lambda{\mathrm{_{res}^{sca}}})$) as a measure of the radiative to dissipative damping for the TM ($E_{\parallel}$) and TE ($E_{\perp}$) polarizations of the incident radiation for the Au, Si and GaAs nanowires (Figures~\ref{rdScat}a1-l1 and a2-l2, respectively). Before, we proceed to analyze the results in Figure~\ref{rdScat}, we note here that the notation `$\lambda{\mathrm{_{res}^{sca}}}$' in the case of parallel polarization for the thin $20\, \mathrm{nm}$ nanowires of the three materials (Au, Si and GaAs) exceptionally refers to the maxima in the scattering efficiency at the ends of the spectrum and the position of the zeroth, first and second-order resonant contributions to the scattering efficiency for the thicker ($r \ge 80\, \mathrm{nm}$) Au nanowires. In all other cases, $\lambda{\mathrm{_{res}^{sca}}}$ refers to the wavelengths at which the scattering resonances occur.

It is universally observed that for nanowires of all three materials (Au, Si, and GaAs), the ratio $\Lambda$ for radiative to dissipative damping for the scattering resonances decreases with increasing temperature (Figure~\ref{rdScat}). This is so even for the cases such as the long-wavelength dielectric resonances outside the absorption band for the semiconductor nanowires wherein an increase in the scattering efficiency $Q_\mathrm{sca}$ is observed at higher temperatures (Figures~\ref{mQscaPar}e, f, h, i, k, l; Figures~\ref{mQscaPer}e, f, h, i, k, l; and, Figures~\ref{rdScat}e1 , e2, f1, f2, h1, h2, i1, i2, k1, k2, l1, l2). The increase in the scattering efficiency at elevated temperatures in these cases is offset by the higher absorption efficiencies at the resonance wavelengths $\lambda{\mathrm{_{res}^{sca}}}$. 

For the TM polarization of the incident radiation in Au nanowires the broad scattering efficiencies exhibit maxima at the edges of the wavelength range under consideration (Figures~\ref{mQscaPar}a-c and Figures~\ref{mQscaPer}a-c). The measure of the radiative to dissipative damping $\Lambda$ shows a decline with temperature due to increased absorption across the entire wavelength range from $400$ to $1450\, \mathrm{nm}$. This decline is greater for the values at longer wavelengths because of a larger relative increase in absorption at the higher temperature of $650 \,\mathrm{K}$ as compared to that for wavelengths close to the localized surface plasmon resonance (LSPR) for Au between $510$ and $600\, \mathrm{nm}$ (Figures~\ref{rdScat}a1, d1, g1, j1). The scattering resonance in Au nanowires for the TE polarization corresponds to the LSPR (Figures~\ref{mQscaPer}a, d, g, j). $\Lambda$ in these cases (Figures~\ref{rdScat}a2, d2, g2, j2), for all nanowire thicknesses considered, is about an order of magnitude smaller than the values observed for the TM polarization (Figures~\ref{rdScat}a1, d1, g1, j1) of the incident radiation because of greater dissipative absorption closer to the LSPR.

For Si nanowires, the $\Lambda$ values for the TE polarization (Figures~\ref{rdScat}b2, e2, h2, k2) are comparable to those for the TM polarization  (Figures~\ref{rdScat}b1, e1, h1, k1) albeit smaller at similar wavelengths. The ratio of radiative to dissipative damping is about $10-15$ orders of magnitude greater for the dielectric resonances of the Si nanowires with $r \ge 80\, \mathrm{nm}$ at long wavelengths outside the absorption band edge where the dissipative damping from absorption is close to zero especially at the lower temperatures (Figures~\ref{rdScat}e1, e2, f1, f2, h1, h2, i1, i2, k1, k2, l1, l2).  Note here the logarithmic $y$-scale for the ratio $\Lambda$ of the radiative to dissipative damping for the TM (Figures~\ref{rdScat}e1, h1, k1) and TE polarizations (Figures~\ref{rdScat}h2, k2). At elevated temperatures, $\Lambda$ declines sharply with an increase in absorption at the longer wavelengths by almost about seven orders of magnitude but is still several orders of magnitude greater than the $\Lambda$ values for Au nanowires due to higher absorption in Au at similar wavelengths. The plasmonic resonances within the absorption band edge at shorter wavelengths for Si nanowires ($r \ge 140\, \mathrm{nm}$) exhibit an order of magnitude decline in $\Lambda$ due to increased dissipative damping at elevated temperatures for TM and TE polarizations alike (Figures~\ref{rdScat}h1, h2, k1, k2). Furthermore, the $80\, \mathrm{nm}$ Si wires are characterized by a complete absence of the long-wavelength dielectric resonances for the TE polarization and exhibit a couple of plasmonic resonances in the scattering efficiency that characteristically deteriorate at higher temperatures due to dissipative damping as reflected in the decreased values for $\Lambda$ (Figures~\ref{rdScat}e1, e2). 

The thin ($20\,\mathrm{nm}$) semiconductor nanowires (Figures~\ref{mQscaPar}b, c and Figures~\ref{mQscaPer}b, c) do not exhibit any scattering or absorption resonances for the two polarizations while the $20\,\mathrm{nm}$ Au wires exhibit an LSPR for the perpendicular polarization alone in the wavelength range $530$ to $630\,\mathrm{nm}$ (Figures~\ref{mQscaPer}a, d, j, g). Despite this, however, the $\Lambda$ values for the semiconductor nanowires are an order of magnitude greater because of low overall dissipative damping as indicated by their low absorption efficiencies (Figures~\ref{rdScat}b1, b2, c1, c2). 

The $\Lambda$ values for the scattering resonances of the GaAs nanowires ($r \ge 80\,\mathrm{nm}$) for the TM and TE polarizations exhibit trends that are similar to those for the Si nanowires (Figures~\ref{rdScat}f1, f2, i1, i2, l1, l2). The $\Lambda$ values for the dielectric resonances for the thicker ($140$ and $200\,\mathrm{nm}$) nanowires, however, are even higher than for those for Si nanowires by about $2-4$ orders of magnitude for both ($T = 200$ and $650\,\mathrm{K}$) (Figures~\ref{rdScat}i1, i2, l1, l2). This is because the absorption efficiencies at these long wavelengths far from the absorption band edge for the direct-bandgap GaAs nanowires are much smaller compared to that for the indirect-bandgap Si nanowires (Figures~\ref{mQabsPar}h, i, k, l and \ref{mQabsPer}h, i, k, l).

\subsubsection*{Absorption resonances}
\begin{figure}[ht!]	
\centering
\includegraphics[scale=0.55]{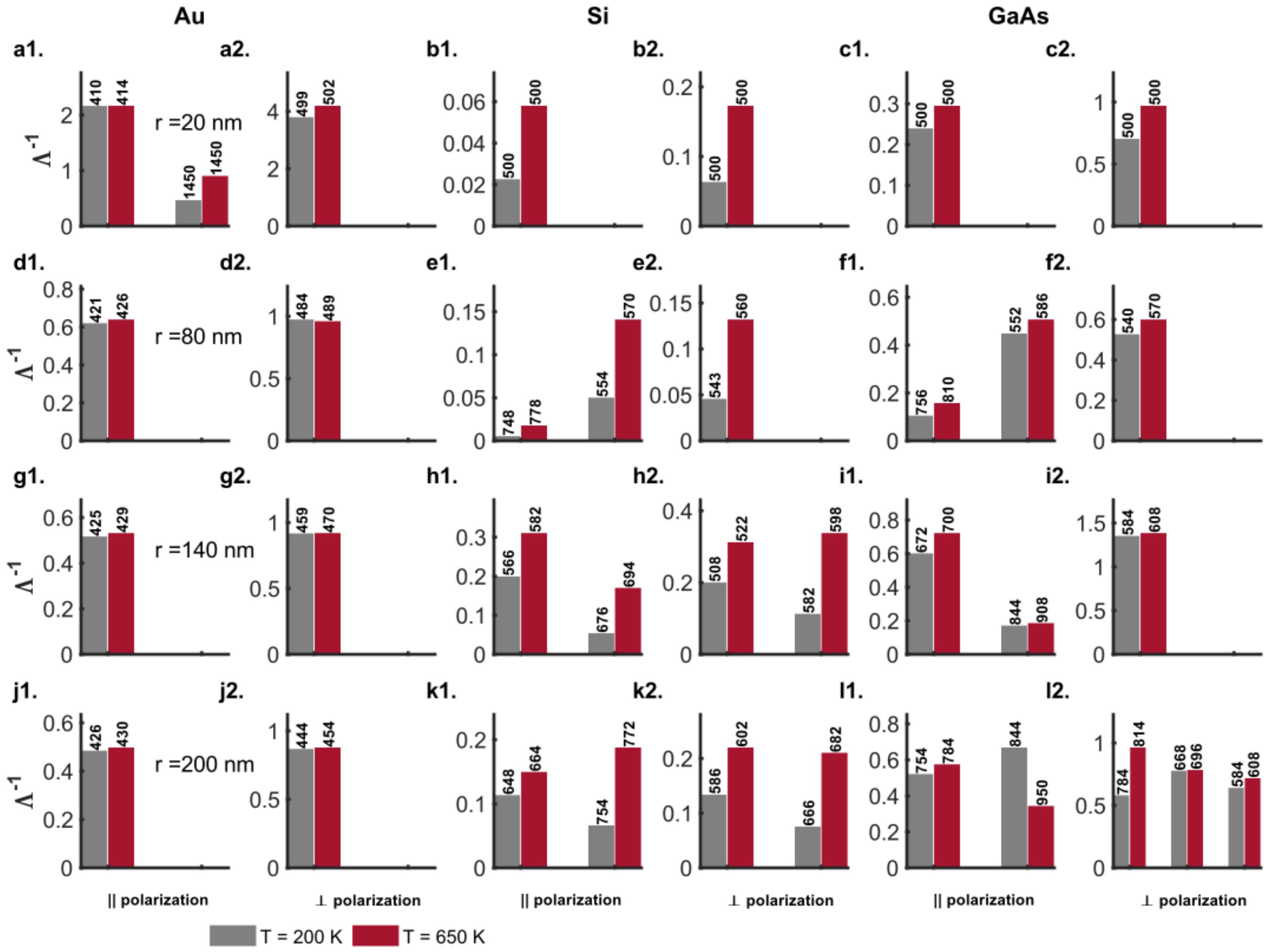} 
\caption{Inverse of the ratio ($\Lambda$) of scattering efficiency ($Q_\mathrm{sca}(\lambda_{\mathrm{res}}^{\mathrm{abs}})$) to the absorption efficiency ($Q_\mathrm{abs}(\lambda_{\mathrm{res}}^{\mathrm{abs}})$) as a measure of the radiative to dissipative damping for the TM ($E_{\parallel}$) and TE ($E_{\perp}$) polarizations of the incident radiation for (a1, d1, g1, j1; a2, d2, g2, j2) Au (DL model), (b1, e1, h1, k1; b2, e2, h2, k2;) Si and (c1, f1, i1, l1; c2, f2, i2, l2) GaAs nanowires of radii $r = 20, 80, 140$ and $200$ nm, respectively, at temperatures $T = 200$ (grey bars) and $650$ (maroon bars) K. The numbers on top of the bars indicate either the wavelengths (in $\mathrm{nm}$) at which the absorption resonances occur or the maxima in absorption efficiency in the wavelength range under consideration in the case of the thin ($20 \,\mathrm{nm}$) wires. Here, the Mie computations for the nanowires of different sizes take into account their thermal expansion, although the text labels indicate the values for nanowire radii at $200$ K.}  
\label{rdAbs} 
\end{figure}

Figure~\ref{rdAbs} shows the ratio ($\Lambda^{-1}$) of resonant absorption efficiency ($Q_\mathrm{abs}(\lambda{\mathrm{_{res}^{abs}}})$) to the scattering efficiency ($Q_\mathrm{sca}(\lambda{\mathrm{_{res}^{abs}}})$) as a measure of the  dissipative absorption to radiative damping for the TM ($E_{\parallel}$) and TE ($E_{\perp}$) polarizations of the incident radiation for the Au, Si and GaAs nanowires (Figures~\ref{rdAbs}a1-l1 and a2-l2, respectively).
The Au nanowires present absorption resonances for the TM (TE) polarization between $410$ and $430\,\mathrm{nm}$ ($440$-$505\,\mathrm{nm}$) for which the $\Lambda^{-1}$ values do not exhibit any significant change with an increase in the temperature from $200$ to $650\,\mathrm{K}$ (Figures~\ref{rdAbs}a1, a2, d1, d2, g1, g2, j1, j2). This implies a proportional and correlated increase in radiative and dissipative damping with an increase in temperature that preserves the ratio $\Lambda^{-1}$. The largest relative increase in $\Lambda^{-1}$ for both TM and TE polarizations with temperature rise between the two semiconductor nanowires is observed for the silicon nanowires (Figures~\ref{rdAbs}b1, b2, e1, e2, h1, h2, k1, k2). This is explained by the indirect-bandgap nature of the Si nanowires characterized by an electron-phonon coupling that strengthens with an increase in temperature, more so than in the case of the direct-bandgap GaAs nanowires. The $\Lambda^{-1}$ values are, however, lower for the Si nanowires when compared to those for Au and GaAs nanowires pointing to stronger radiation damping through scattering associated with them (Figure~\ref{rdAbs}).  
Among the two polarizations for the incident radiation, the Si and GaAs semiconductor nanowires present larger values of $\Lambda^{-1}$ for the TE polarization as a result of lower scattering efficiencies or radiation damping associated with the TE polarization (Figures~\ref{rdAbs}b1, b2, c1, c2, e1, e2, f1, f2, h1, h2, i1, i2, k1, k2, l1, l2 and Figures~\ref{mQscaPer}b, c, e, f, h, i, k, l).

\section*{Summary and Conclusions}

We have studied the effects of temperature on the Mie resonances for the polarization-dependent scattering and absorption efficiencies of metallic (Au), and indirect (Si) and direct (GaAs) bandgap semiconductor nanowires. Similar to nanoparticles \cite{RN159}, metallic Au nanowires also exhibit broad absorption resonances that extend far into the long-wavelength NIR regime and tend to increase with increasing temperature. This is a key characteristic of the Au nanostructures that makes them susceptible to physicochemical deterioration at elevated temperatures. 

Scattering resonances in the case of the TE polarization of the incident EM radiation for the nanowires occur at shorter wavelengths as a consequence of 
confinement 
along the radial direction of the cylindrical nanowires. Plasmonic resonances that occur within the absorption-band edge of the semiconductor nanowires at shorter wavelengths and those for the Au nanowires are observed to deteriorate at elevated temperatures regardless of the polarization of the incident EM wave. In contrast the phononic contribution to the dielectric resonances in semiconductor nanowires that occur away from the absorption band, however, edge serves to strengthen them at elevated temperatures. This augurs well for applications in thermophotovoltaics and high temperature insulators wherein the broadband reflectances shall actually increase with rising temperatures and result in enhanced performance at elevated temperatures. The thin ($\mathrm{20\,nm}$) Si nanowires offer a low absorption efficiency for both the TE and TM polarizations of the incident radiation in conjunction with high scattering efficiency overall. However, this contrasts with the high absorption efficiencies and comparable (to Si) scattering efficiency 
of the GaAs nanowires of the same radius. This implies that thin indirect-bandgap semiconductor nanowires may be better suited, compared to the direct-bandgap nanowires, for use as highly efficient scatterers in the case of unpolarized incident radiation or with the electric field polarized along the cylinder axis (TM mode). Typically, for obtaining very large scattering efficiencies multilayered cylindrical geometries are required to engineer coincident multimode resonances \cite{RN206}. Here, in our results for the thin Si nanowires, it is just the zeroth order TM mode $b_0$ that contributes to the high scattering efficiency. 

The results also indicate the presence of temperature-resilient higher-order anapole modes in thin Si and GaAs nanowires ($r=125$-$130 \,\mathrm{nm}$) characterized by low scattering and high absorption efficiencies for the TE polarization in the visible wavelength range between $570$-$610 \,\mathrm{nm}$. Elevated temperatures for these TE anapole modes present even higher absorption efficiencies in this range with values for GaAs nanowires roughly $3-7$ times greater than those for the indirect-bandgap Si nanowires. This makes the direct-bandgap GaAs nanowires ideal for temperature-resilient applications such as scanning near field microscopy (SNOM), cloaking, localized heating, nonivasive sensing or detection that seek to exploit strong localization of energy in the near field. 

The relative strength of the radiative to dissipative damping was studied for the TM and TE polarizations of the incident EM radiation for Au, Si and GaAs nanowires as a function of the temperature by examining the ratio of scattering to absorption efficiency and its inverse for the scattering and absorption resonances, respectively. 
Extremely large values for the ratio of the  resonant scattering efficiency to the absorption efficiency ($\Lambda$), that decrease at elevated temperatures, are observed for dielectric resonances away from the absorption band edge at long wavelengths for both polarizations in semiconductor nanowires. These large values of $\Lambda$ can be attributed to the low absorption efficiencies of the two semiconductors at these wavelengths. The ratio of radiative to dissipative damping remains unchanged for the Au nanowires regardless of polarization with an increase in temperature as reflected in comparable values for $\Lambda$. The Si nanowires register the largest increase in $\Lambda^{-1}$ for absorption resonances with an increase in temperature. However, these values for $\Lambda^{-1}$ are lower than those for the Au and GaAs nanowires on account of stronger radiation damping.

\section*{Competing interests}

The authors declare no competing interests.

\section*{Author contributions}

All authors contributed to the design, implementation and analysis of the research, interpretation of the results, and writing of the manuscript. All authors have given their approval for the final
version of the manuscript.

\section{Data availability}
The data is available from authors upon reasonable request.

\section*{Supporting Information}
The Supporting Information includes four additional figures S1-S4 that show the scattering and absorption efficiencies for Au, Si and GaAs as a function of the nanowire radius $r$ and wavelength $\lambda$ of the incident EM radiation at temperatures $T = 200$, $470$, and $650 \,\mathrm{K}$ for the TM and TE polarizations.

\section*{Acknowledgments}
The authors
acknowledge funding and support from the Academy of Finland, COMP Center of Excellence Programs (2015-2017), Grant No.~284621; Quantum Technology Finland Center of Excellence Program, Grant No.~312298; Radiation Detectors for Health, Safety and Security (RADDESS) Consortium Grant of the Academy of Finland; the Aalto University Energy Efficiency Research Program (EXPECTS); the Aalto Science-IT project; the Discovery Grants Program of  the Natural Sciences and Engineering Research Council (NSERC) of Canada, and Canada Research Chairs Program. Computational resources were provided by Compute Canada (www.computecanada.ca).


\providecommand{\latin}[1]{#1}
\makeatletter
\providecommand{\doi}
  {\begingroup\let\do\@makeother\dospecials
  \catcode`\{=1 \catcode`\}=2 \doi@aux}
\providecommand{\doi@aux}[1]{\endgroup\texttt{#1}}
\makeatother
\providecommand*\mcitethebibliography{\thebibliography}
\csname @ifundefined\endcsname{endmcitethebibliography}
  {\let\endmcitethebibliography\endthebibliography}{}

\end{document}


\maketitle

\begin{figure}[ht!]	
\centering
\includegraphics[scale=0.64]{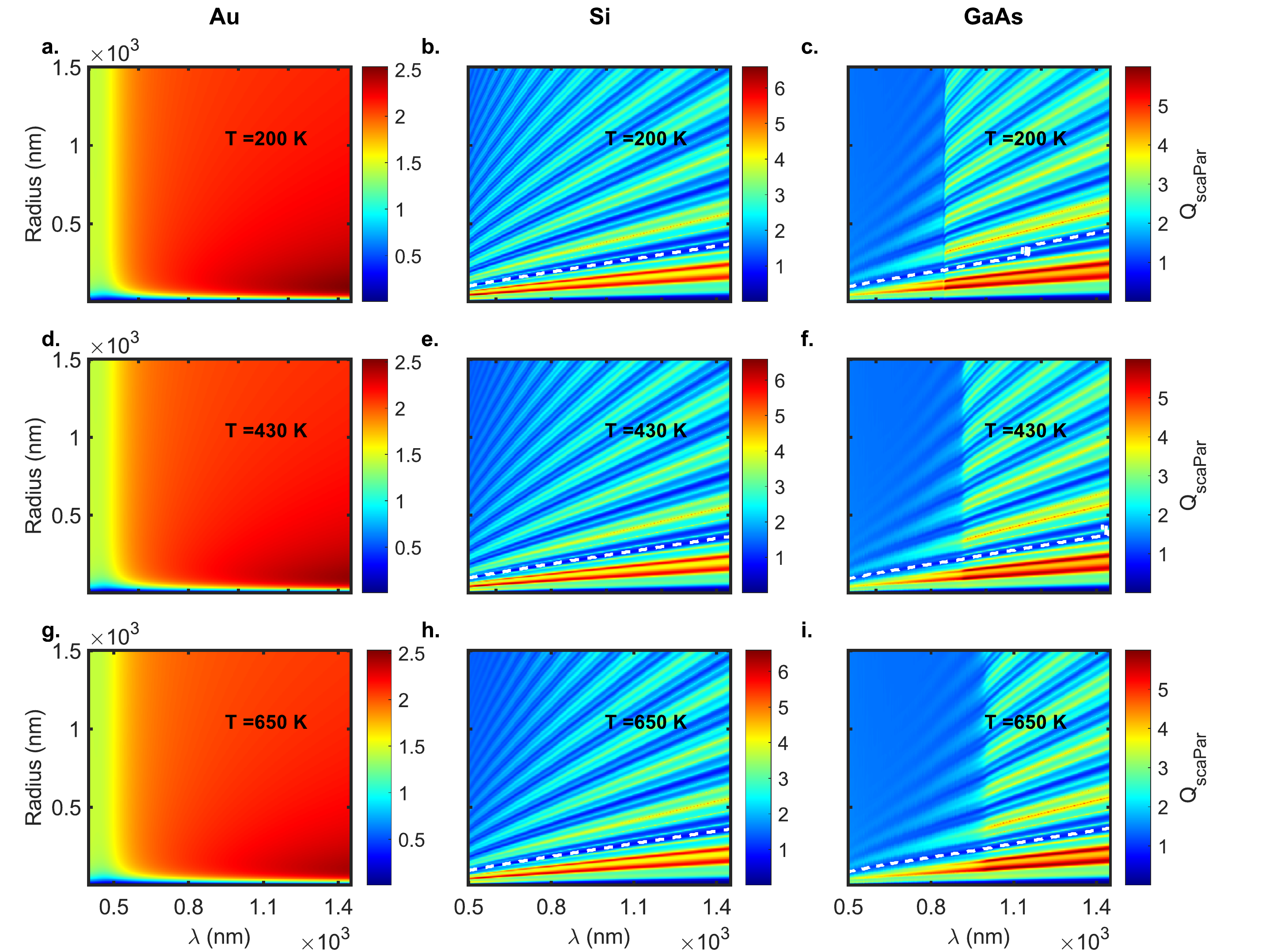} 
\caption{Mie scattering efficiency, $Q_\mathrm{scaPar}$, for the TM ($E_{\parallel}$) polarization of the incident radiation as a function of the wavelength, $\lambda$, of the incident radiation and the nanowire radii $r$ at three different temperatures: $T=200$ (a-c), $430$ (d-f) and $650$ K (g-i) for Au (DL model), Si and GaAs nanowires, respectively. (See Figure 1 also.) The white dashed lines in sub-plots for Si (b, e, h) and GaAs (c, f, i) nanowires show the trajectory of the minima in $Q_\mathrm{scaPar}$ as a function of the nanowire radii ($r > 50 \,\mathrm{nm}$) and wavelength $\lambda$ of the incident radiation.}  
\label{figs-QscaPar2d} 
\end{figure}

\begin{figure}[ht!]	
\centering
\includegraphics[scale=0.64]{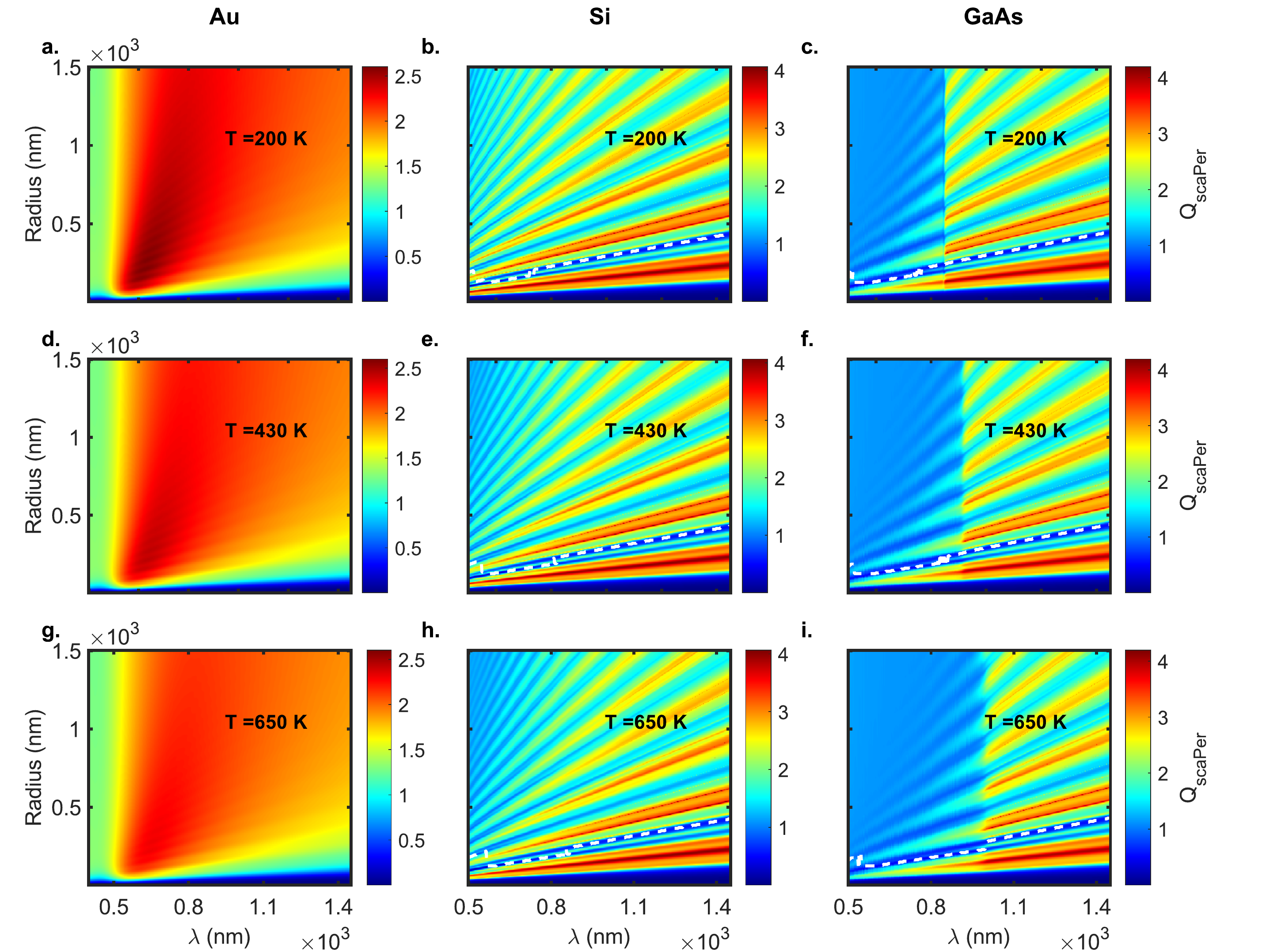} 
\caption{Mie scattering efficiency, $Q_\mathrm{scaPer}$, for the TE ($E_{\perp}$) polarization of the incident radiation as a function of the wavelength, $\lambda$, of the incident radiation and the nanowire radii $r$ at three different temperatures: $T=200$ (a-c), $430$ (d-f) and $650$ K (g-i) for Au (DL model), Si and GaAs nanowires, respectively. (See Figure 2 also.) The white dashed lines in sub-plots for Si (b, e, h) and GaAs (c, f, i) nanowires show the trajectory of the minima in $Q_\mathrm{scaPer}$ as a function of the nanowire radii ($r > 125 \,\mathrm{nm}$) and wavelength $\lambda$ of the incident radiation.}  
\label{figs-QscaPer2d} 
\end{figure}

\begin{figure}[ht!]	
\centering
\includegraphics[scale=0.64]{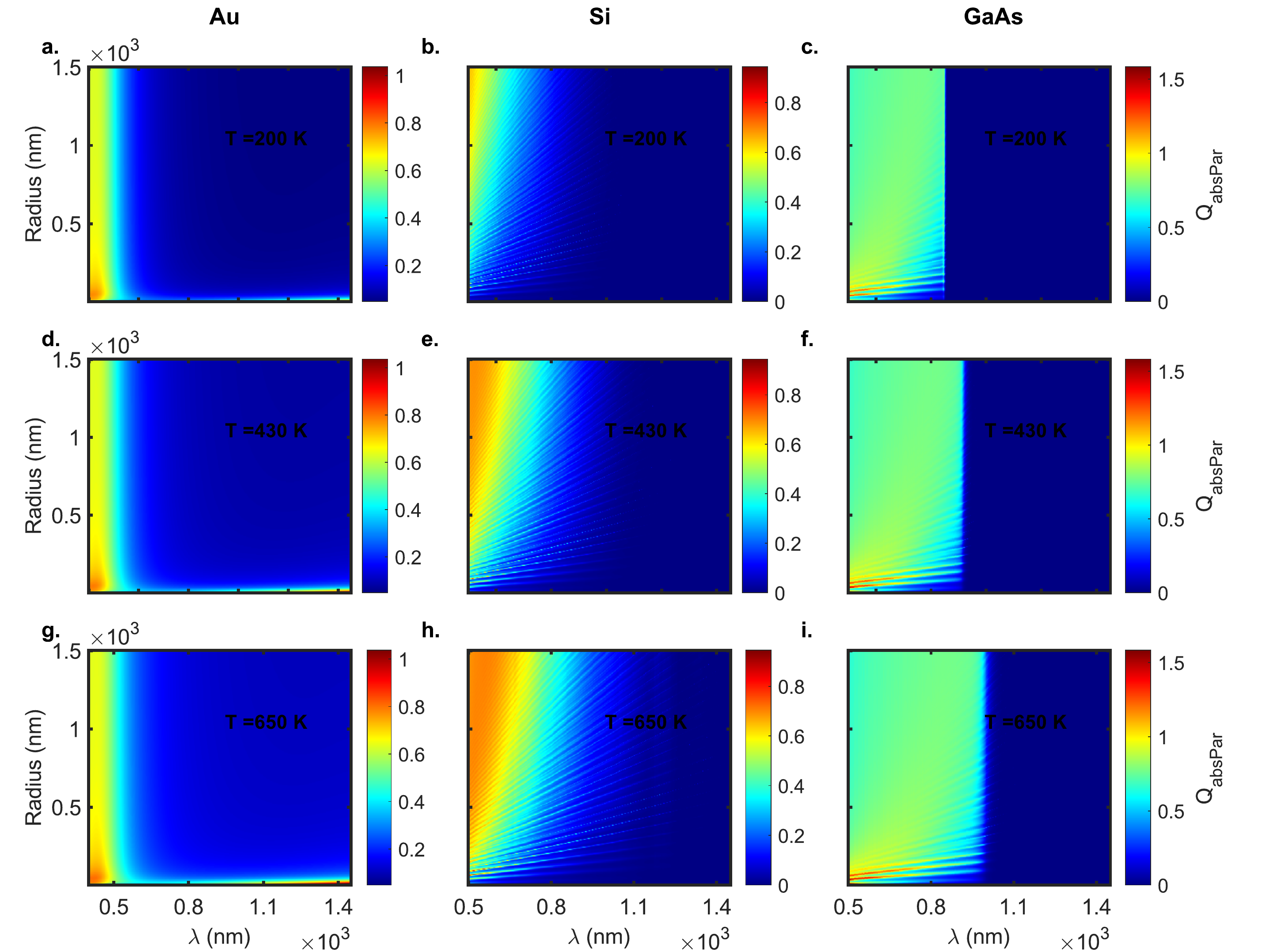} 
\caption{Mie absorption efficiency, $Q_\mathrm{absPar}$, for the TM ($E_{\parallel}$)  polarization of the incident radiation as a function of the wavelength, $\lambda$, of the incident radiation and the nanowire radii $r$ at three different temperatures $T=200$ (a-c), $430$ (d-f) and $650$ K (g-i) for Au (DL model), Si and GaAs nanowires, respectively. (See Figure 3 also.)} 
\label{figs-QabsPar2d} 
\end{figure}

\begin{figure}[ht!]	
\centering
\includegraphics[scale=0.64]{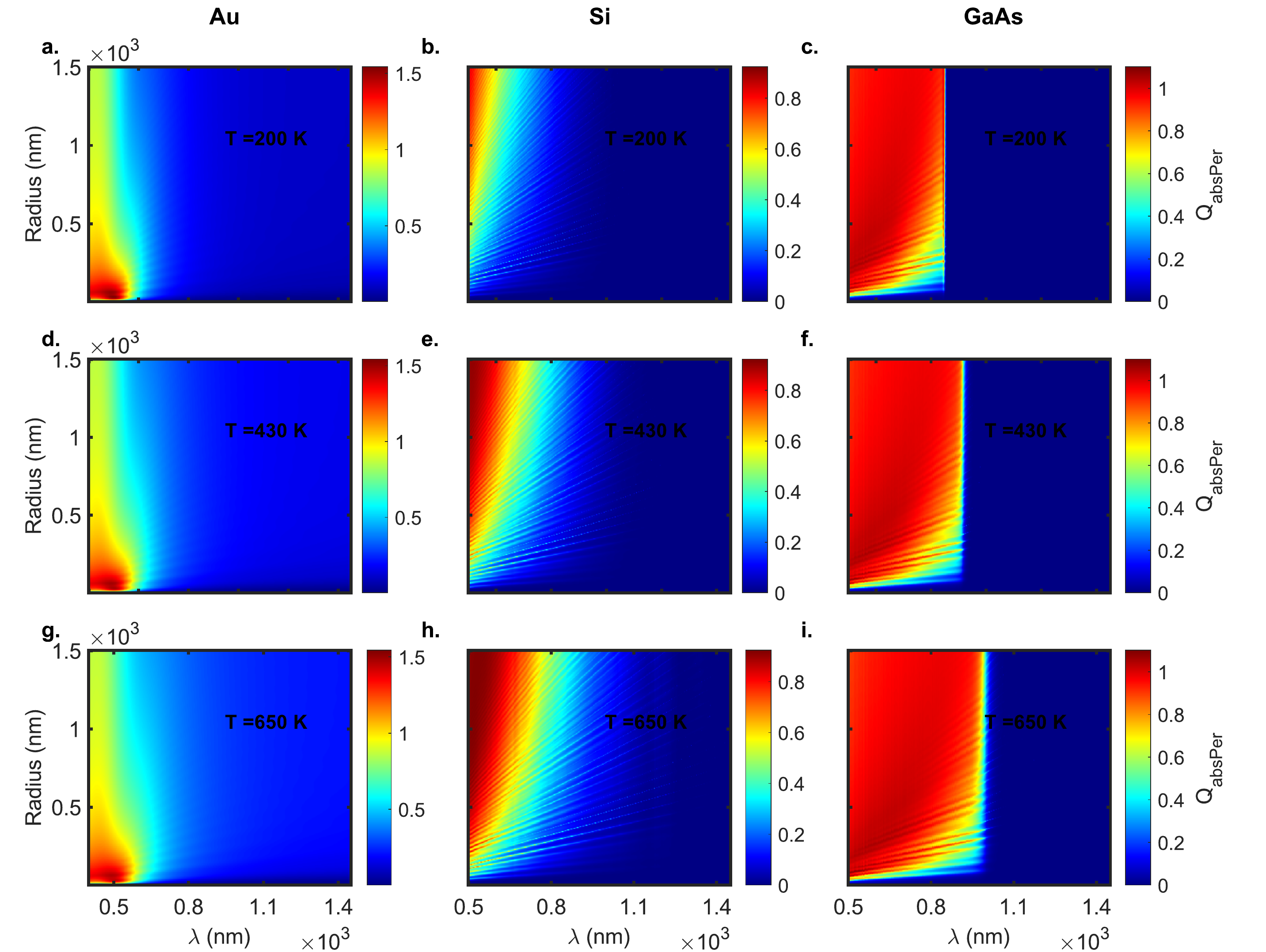} 
\caption{Mie absorption efficiency, $Q_\mathrm{absPer}$, for the TE ($E_{\perp}$) polarization of the incident radiation as a function of the wavelength, $\lambda$, of the incident radiation and the nanowire radii $r$ at three different temperatures $T=200$ (a-c), $430$ (d-f) and $650$ K (g-i) for Au (DL model), Si and GaAs nanowires, respectively.  (See Figure 4 also.)} 
\label{figs-QabsPer2d} 
\end{figure}

